\documentclass[fleqn,usenatbib]{mnras}

\usepackage{newtxtext,newtxmath}

\usepackage[T1]{fontenc}

\DeclareRobustCommand{\VAN}[3]{#2}
\let\VANthebibliography\thebibliography
\def\thebibliography{\DeclareRobustCommand{\VAN}[3]{##3}\VANthebibliography}

\usepackage{graphicx}	
\usepackage{amsmath}	
\usepackage{orcidlink}

\usepackage{tablefootnote}
\usepackage[flushleft]{threeparttable}
\usepackage{booktabs}
\usepackage{siunitx}
\usepackage{tabularx}
\usepackage{xspace}
\usepackage{pdflscape}
\usepackage{multirow} 
\usepackage{enumitem}   
\usepackage{chemformula}
\usepackage{afterpage}
\usepackage{multicol}
\usepackage{placeins}
\usepackage{hyperref}
\usepackage{subfigure}

\newcommand{\mystar}{TrES-4\xspace}
\newcommand{\planet}{TrES-4\,b\xspace}
\newcommand{\tiberius}[1]{\texttt{Tiberius}\xspace}
\newcommand{\eureka}[1]{\texttt{Eureka!}\xspace}
\newcommand{\poseidon}[1]{\texttt{POSEIDON}\xspace}
\newcommand{\prt}{\texttt{petitRADTRANS}\xspace}

\newcommand{\vdag}{\dagger}

\defcitealias{sozzettiGAPSProgrammeHARPSN2015}{S15}
\defcitealias{penzlinBOWIEALIGNHowFormation2024}{Penzlin \& Booth et al. (2024)}
\defcitealias{moranHighTideRiptide2023}{Moran \& Stevenson et al. (2023)}
\defcitealias{thejwsttransitingexoplanetcommunityearlyreleasescienceteamIdentificationCarbonDioxide2023}{JWST Transiting Exoplanet Community Early Release Science Team 2023}
\defcitealias{ahrerEarlyReleaseScience2023}{Ahrer et al. 2023}
\defcitealias{feinsteinEarlyReleaseScience2023}{Feinstein et al. 2023}
\defcitealias{aldersonEarlyReleaseScience2023}{Alderson et al. 2023}
\defcitealias{rustamkulovEarlyReleaseScience2023}{Rustamkulov et al. 2023}
\defcitealias{Madhusudhan2014}{Madhusudhan et al. 2014}

\title[BOWIE-ALIGN: the aligned TrES-4b]{BOWIE-ALIGN: Sub-stellar metallicity and carbon depletion in the aligned TrES-4b with JWST NIRSpec transmission spectroscopy}

\author[A. Meech et al.]{Annabella Meech $^{\orcidlink{0000-0002-7500-7173},1}$\thanks{E-mail: annabella.meech@cfa.harvard.edu},
Alastair B. Claringbold $^{\orcidlink{0000-0003-1309-5558},2,3}$,
Eva-Maria Ahrer $^{\orcidlink{0000-0003-0973-8426},4}$,
James Kirk $^{\orcidlink{0000-0002-4207-6615},5}$,\newauthor
Mercedes L\'opez-Morales $^{\orcidlink{0000-0003-3204-8183},6}$,
Jake Taylor $^{\orcidlink{0000-0003-4844-9838},7}$,
Richard~A. Booth $^{\orcidlink{0000-0002-0364-937X},8}$,
Anna~B.T. Penzlin $^{\orcidlink{0000-0002-8873-6826},5}$,
\newauthor
Lili Alderson $^{\orcidlink{0000-0001-8703-7751},9}$,
Duncan A. Christie $^{\orcidlink{0000-0002-4997-0847},4}$,
Emma Esparza-Borges $^{\orcidlink{0000-0002-2341-3233},10,11}$,
Charlotte Fairman $^{\orcidlink{0000-0001-9665-5260},12}$,\newauthor
Nathan J. Mayne $^{\orcidlink{0000-0001-6707-4563},13}$,
Mason McCormack $^{\orcidlink{0000-0002-1463-9847},14}$,
James E. Owen $^{\orcidlink{0000-0002-4856-7837},5}$,
Vatsal Panwar $^{\orcidlink{0000-0002-2513-4465}, 2,3}$,\newauthor
Diana Powell $^{\orcidlink{0000-0002-4250-0957},14}$,
Denis E. Sergeev $^{\orcidlink{0000-0001-8832-5288},12}$,
Daniel Valentine $^{\orcidlink{0000-0002-2643-6836},12}$,
Hannah R. Wakeford $^{\orcidlink{0000-0003-4328-3867},12}$,\newauthor
Peter J.\ Wheatley $^{\orcidlink{0000-0003-1452-2240},2,3}$,
Maria Zamyatina $^{\orcidlink{0000-0002-9705-0535},13}$
\\
$^{1}$Center for Astrophysics | Harvard \& Smithsonian, 60 Garden St, Cambridge, MA 02138, USA\\
$^{2}$Centre for Exoplanets and Habitability, University of Warwick, Gibbet Hill Road, Coventry CV4 7AL, UK\\
$^{3}$Department of Physics, University of Warwick, Gibbet Hill Road, Coventry CV4 7AL, UK\\
$^{4}$ Max-Planck-Institut f\"{u}r Astronomie, K\"{o}nigstuhl 17, 69117 Heidelberg, Germany\\
$^{5}$Department of Physics, Imperial College London, Prince Consort Rd, London, SW7 2AZ, UK \\
$^{6}$Space Telescope Science Institute, 3700 San Martin Drive, Baltimore, MD 21218, USA\\
$^{7}$Department of Physics, University of Oxford, Parks Rd, Oxford, OX1 3PU, UK\\
$^{8}$School of Physics and Astronomy, University of Leeds, Leeds LS2 9JT, UK\\
$^{9}$Department of Astronomy, Cornell University, 122 Sciences Drive, Ithaca, NY 14853, USA\\
$^{10}$Instituto de Astrof\'isica de Canarias, C. V\'ia L\'actea, San Crist\'obal de La Laguna 38205, Spain\\
$^{11}$Department of Astrophysics, University of La Laguna, San Crist\'obal de La Laguna 38200, Spain\\
$^{12}$School of Physics, University of Bristol, HH Wills Physics Laboratory, Tyndall Avenue, Bristol BS8 1TL, UK\\
$^{13}$ Department of Physics and Astronomy, University of Exeter, Exeter, UK\\
$^{14}$Department of Astronomy \& Astrophysics, University of Chicago, Chicago, IL 60637, USA\\
}

\date{Accepted 2025 March 26. Received 2025 March 21; in original form 2025 January 15}
\pubyear{\the\year{}}

\begin{document}
\label{firstpage}
\pagerange{\pageref{firstpage}--\pageref{lastpage}}
\maketitle

\begin{abstract}
The formation and migration history of a planet is expected to be imprinted in its atmosphere, in particular its carbon-to-oxygen (C/O) ratio and metallicity. 
The BOWIE-ALIGN programme is performing a comparative study of JWST spectra of four aligned and four misaligned hot Jupiters, with the aim of characterising their atmospheres and corroborating the link between the observables and the formation history.
In this work, we present the $2.8-5.2$\,micron transmission spectrum of \planet, a hot Jupiter with an orbit aligned with the rotation axis of its F-type host star.
Using free chemistry atmospheric retrievals, we report a confident detection of H$_2$O at an abundance of $\log X_\mathrm{H_2O}=-2.98^{+0.68}_{-0.73}$ at a significance of $8.4\sigma$.
We also find evidence for CO and small amounts of CO$_2$, retrieving abundances $\log X_\mathrm{CO}= -3.76^{+0.89}_{-1.01}$ and $\log X_\mathrm{CO_2}= -6.86^{+0.62}_{-0.65}$ ($3.1\sigma$ and $4.0\sigma$ respectively).
The observations are consistent with the the atmosphere being in chemical equilibrium; our retrievals yield $\mathrm{C/O}$ between $0.30-0.42$ and constrain the atmospheric metallicity to the range $0.4-0.7\times$ solar.
The inferred sub-stellar properties (C/O and metallicity) challenge traditional models, and could have arisen from an oxygen-rich gas accretion scenario, or a combination of low-metallicity gas and carbon-poor solid accretion.

\end{abstract}
%




\section{Introduction}

In the field of exoplanet atmospheres, we are in the era of exquisite data sensitivity, sufficient to probe chemical abundances and derive precise carbon-to-oxygen ($\mathrm{C/O}$) ratios and atmospheric metallicities, thanks to the resolution and near-infrared wavelength coverage of the instruments on JWST \citepalias[][]{thejwsttransitingexoplanetcommunityearlyreleasescienceteamIdentificationCarbonDioxide2023,rustamkulovEarlyReleaseScience2023,feinsteinEarlyReleaseScience2023,aldersonEarlyReleaseScience2023,ahrerEarlyReleaseScience2023}.
The challenge is now to interpret these measurements, linking them to exoplanet formation and evolution history.
Early ideas \citep[e.g.,][]{obergEffectsSnowlinesPlanetary2011,madhusudhanRatioDimensionCharacterizing2012} suggested that atmospheric $\mathrm{C/O}$ can be used to inform where a planet formed in the protoplanetary disc with respect to snow lines of the dominant volatiles, such as \ch{H2O}, \ch{CO} and \ch{CO2}. 
However, more recent studies have challenged that idea. 
For example, the radial composition profile of the disc varies between systems \citep{lawMoleculesALMAPlanetforming2021}, with different initial metallicities and host stars.
Also, the radial location of snowlines in protoplanetary discs evolve over time \citep[][]{morbidelliFossilizedCondensationLines2016,owenSnowlinesCanBe2020,2018Eistrup}, and volatile-carrying solids drift through the discs \citep[e.g.,][]{Booth2017,2021Schneider}.
More recently, \citetalias{penzlinBOWIEALIGNHowFormation2024} demonstrated that the highly unconstrained nature of key planet formation and disc parameters -- e.g.,  temperature profile, dust-to-gas ratio and chemical composition within the disc -- make it difficult to precisely predict the atmospheric composition of individual exoplanets from formation models. 
However, the key insight from \citetalias{penzlinBOWIEALIGNHowFormation2024} is that comparing \textit{populations} of planets (as opposed to single systems) with different migration histories could constrain planet formation models.
Specifically, they show that the C/O and metallicity of close-in exoplanets that evolve through migration in a disc should be different from close-in exoplanets that undergo disc-free migration, after formation at another location in the disc. 

{%
\renewcommand{\arraystretch}{1.2}
\begin{table}
\caption{Adopted system parameters.}

    \begin{threeparttable}
        
    \resizebox{0.8\columnwidth}{!}{
    \begin{tabular}{lc}
    \toprule
         Parameter [Unit] & Value  \\
         \midrule
         \textbf{Star} \\
         $T_*$\,[K] & $6295\pm65$  \\
         $M_*\,[M_\odot]$ & $1.45\pm0.05$ \\
         $R_*\,[R_\odot]$ & $1.81\pm0.08$ \\
         $\log g_*$\,[cgs] & $4.09\pm0.03$ \\
         $\mathrm{[Fe/H]}$ & $0.28\pm0.09$ \\
         \midrule
         \textbf{Planet} \\
         $P$\,[days] & $3.55392889\pm0.00000044^\vdag$ \\
         $M_\mathrm{p} [M_\mathrm{J}]$ & $0.494\pm0.035$\\
         $R_\mathrm{p} [R_\mathrm{J}]$ &$1.838{\substack{+0.081\\-0.090}}$ \\
         $T_\mathrm{eq}$ [K] & $1795{\substack{+35\\-39}}$\\
         $e\cos\omega$ & $0.0010{\substack{+0.0022\\-0.0017}}$\\
         $e\sin\omega$ & $0{\substack{+0.012\\-0.022}}$ \\

    \bottomrule
    \end{tabular}
    }
    \begin{tablenotes}
        \item[$\vdag$] All values from \citet{sozzettiGAPSProgrammeHARPSN2015}, with the exception of the period, $P$, from \citet{kokoriExoClockProjectIII2023}.
    \end{tablenotes}
    \end{threeparttable}
    \label{table:system_params}
\end{table}
}

The BOWIE-ALIGN survey (Bristol, Oxford, Warwick, Imperial, Exeter - A spectral Light Investigation into gas Giant origiNs; JWST GO 3838; PIs: Kirk \& Ahrer) seeks to characterise the atmospheres of eight hot Jupiters in order to test tracers of planet formation.
The prudently curated sample consists of four planets believed to have migrated through a disc, and four believed to have undergone disc-free migration, based on the current alignments of their orbital planes with respect to the rotational plane of their host stars.
The targets are considered `aligned' if their sky-projected obliquity $|\lambda|<30^\circ$, and `misaligned' if $|\lambda|>45^\circ$, based on the definition in \citet{spaldingTidalErasureStellar2022}.
The main objectives, methods, and target selection for this programme are outlined in the survey paper, \citet{kirkBOWIEALIGNJWSTComparative2024}, while the theoretical basis is explored in \citetalias{penzlinBOWIEALIGNHowFormation2024}.
To date, the transmission spectrum of one target from this program has been published, the misaligned hot Jupiter WASP-15\,b.
In \citet{kirkBOWIEALIGNJWSTReveals2025}, we found WASP-15\,b to host a super-stellar metallicity atmosphere, with a solar C/O, and evidence of \ch{SO2} absorption, the combination of which points to late planetesimal accretion. 
Here we present the transmission spectrum of the second target in the BOWIE-ALIGN sample, \planet.

The hot Jupiter \planet \citep[$R_\mathrm{p}=1.838{\substack{+0.081\\-0.090}}\,R_\mathrm{J}$, $M_\mathrm{p}=0.494\pm0.035\,M_\mathrm{J}$, $T_\mathrm{eq}=1795$\,K;][]{sozzettiGAPSProgrammeHARPSN2015}, discovered by \citet{mandushevTrES4TransitingHot2007}, is one of the four aligned planets in the sample.
It orbits an F-type host star \citep[][]{mandushevTrES4TransitingHot2007}, which is crucially above the Kraft break, as are all of the targets in the programme.
The `Kraft break', at $T_\mathrm{eff}\simeq 6100$\,K, is a distinct shift in the rotation rates of stars, thought to occur due to the magnetic braking contribution of the convective zone in cooler stars \citep[][]{kraftStudiesStellarRotation1967,beyerKraftBreakSharply2024}.
It was later observed that the hotter stars above the Kraft break are seen to host hot Jupiters with various orbital obliquities, while the cooler stars below show predominantly low stellar obliquities \citep{winnHotStarsHot2010,albrechtOBLIQUITIESHOTJUPITER2012}.
Regarding the latter, misaligned hot Jupiters are thought to invoke stellar tides which can act to realign the stellar spin axis with the planet orbital plane.
This is more efficient for the cooler stars; with deeper surface convective zones and consequent stronger magnetic braking, these stars have lower angular momentum which can be overcome by the planet \citep[][]{linTidalInteractionsSpinorbit2017,dawsonTidalOriginHot2014}.
On the reverse, stars above the Kraft break have thin surface convective zones, which do not produce efficient tides. 
There is a period in which stars above the Kraft break (with masses $1.2\,M_\odot<M_*\lesssim5\,M_\odot$\footnote{Stars more massive than $4-5$~M$_\odot$ become radiative on the pre-main-sequence \citep[e.g.][]{Palla1993}, and are often still accreting when they reach the zero-age main-sequence \citep{Zinnecker2007}.}) may tidally re-align, since they are fully convective during their pre-main-sequence phase \citep[][]{spaldingTidalErasureStellar2022}.
However, \planet\ would need to orbit within 0.02\,AU of the star to have tidally re-aligned within the $\sim 20$~Myr period during which its host star was convective \citep{spaldingTidalErasureStellar2022}. 
Since the \planet lies at 0.05\,AU, it is highly unlikely to have undergone tidal re-alignment \emph{post} migration.
In this work, we base conclusions upon the assumption that \planet exhibits primordial alignment; its obliquity was measured by \citet{naritaSpinOrbitAlignmentTrES42010} to be $\lambda=6.3\pm4.7^{\circ}$.
As outlined in \citet{kirkBOWIEALIGNJWSTComparative2024}, the presumption is thus that \planet formed in the outer regions of its system and consequently migrated towards the star, through the disc.

In this work, we aim to characterise the atmosphere of this hot Jupiter.
We focus on constraining its C/O and metallicity, towards the wider goal of identifying trends in planet formation scenarios within the wider context of the BOWIE-ALIGN survey.
We provide a description of the observations in \S\ref{SECTION-2:observations}.
The procedures used to reduce the data and the results, including the transmission spectrum, are presented in \S\ref{SECTION-3:data-reduction}.
We then present our retrieval analysis of the extracted transmission spectrum in \S\ref{SECTION-4:retrievals}.
Finally, the discussion of our results and concluding remarks are outlined in \S\ref{SECTION-5:discussion} and \S\ref{SECTION-6:conclusions}.

\section{Observations}

\begin{figure*}
\caption{\textbf{Left:} Extracted white light curves using the \texttt{Tiberius} pipeline, with best-fit models overplotted in black. 
These have been plotted against time, having subtracted the best-fitting mid-transit time. \textbf{Right:} Residuals and associated histograms between the light curves and their best-fit models. 
The in-transit integrations are shaded in purple for reference.}
\centering
\includegraphics[width=0.94\linewidth]{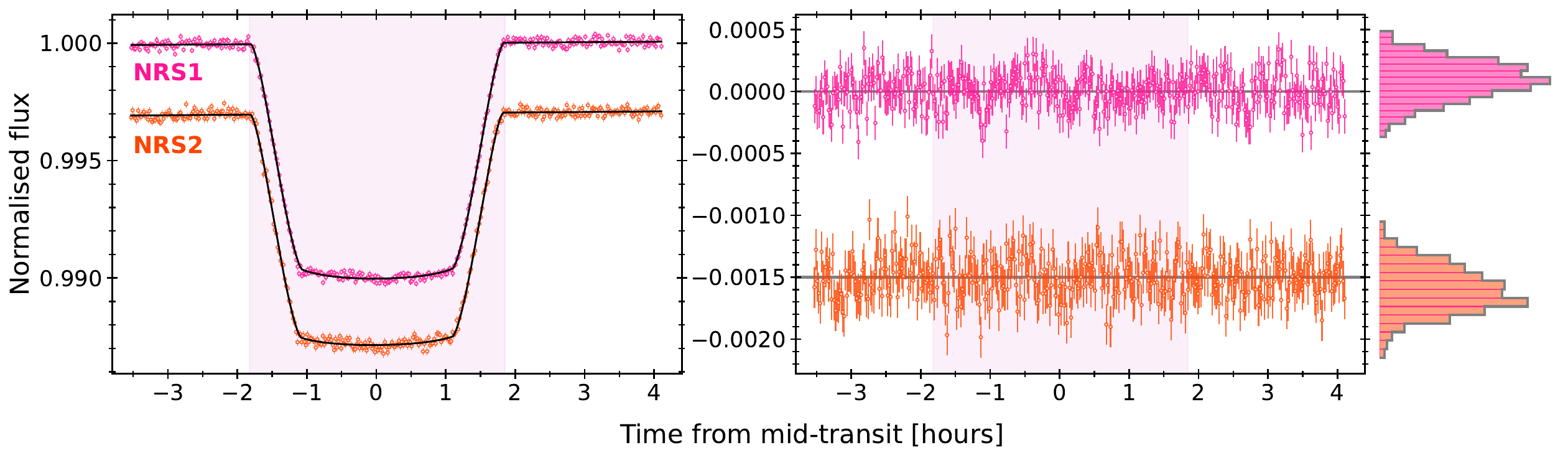}
\label{fig:WLCs}
\end{figure*}

\label{SECTION-2:observations}
\mystar was observed by JWST as part of the GO 3838 programme.
We observed the full transit of planet b, using the NIRSpec instrument \citep{jakobsenNearInfraredSpectrographNIRSpec2022} in Bright Object Time Series (BOTS) mode, with the NRSRAPID readout pattern.
We used the G395H grating, F290LP filter, the S1600A1 aperture and SUB2048 subarray.
This setting provides a wavelength coverage of $2.8-5.2$\,microns, and an average resolution of $R= 2700$.
There exists a physical gap between the NIRSpec detectors. 
The positioning of this gap in wavelength depends on the dispersion-filter combination used, and falls between $3.72-3.82$\,microns for G395H/F290LP.
For the target acquisition, we used the WATA mode of NIRSpec on a fainter, nearby star 2MASSJ17531241+3712396, with the SUB32 array and S1600A1 slit.
The science observation started at 01:27:06.022 UTC on 17th June 2024, and consisted of a total 378 integrations over 7.65\,hrs, which included 180 integrations (3.65\,hr) in-transit, 85 pre- and 113 post-transit baseline (1.73 and 2.27\,hrs respectively), with 80 groups per integration.

\section{Data reduction}

\label{SECTION-3:data-reduction}
To reduce these data, we used two separate pipelines, \tiberius\ \footnote{\url{https://tiberius.readthedocs.io/en/latest/}} \citep{kirkRayleighScatteringTransmission2017,kirkACCESSLRGBEASTSPrecise2021} and \eureka\ \footnote{\url{https://eurekadocs.readthedocs.io/en/latest/}} \citep[v.1.1;][]{bellEurekaEndEndPipeline2022}.
We detail each in turn below.
Multiple, independent reductions have been standard in the community, as recommended by the work of the Early Release Science programme \citepalias[][]{thejwsttransitingexoplanetcommunityearlyreleasescienceteamIdentificationCarbonDioxide2023,ahrerEarlyReleaseScience2023,aldersonEarlyReleaseScience2023,feinsteinEarlyReleaseScience2023,rustamkulovEarlyReleaseScience2023}, and allow us to check how robust the transmission spectrum is to reduction choices.
Additionally, we present a second independent reduction using the \tiberius\ pipeline in Appendix~\ref{SECTION-7:appendix-tiberiusJK}.

\subsection{\texttt{Tiberius}}
\label{section:tiberius-LCs}
\tiberius\ is an open-source \texttt{Python} package, written to reduce and fit exoplanet transit light curves \citep[][]{kirkRayleighScatteringTransmission2017,kirkACCESSLRGBEASTSPrecise2021}.
The pipeline has been used by the community to reduce JWST exoplanet time-series data \citep[see e.g.,][]{moranHighTideRiptide2023,esparza-borgesDetectionCarbonMonoxide2023,kirkJWSTNIRCamTransmission2024}.
Our implementation, described below, largely follows the standard routine, and is similar to that described in \citet{kirkBOWIEALIGNJWSTReveals2025}.

\subsubsection{Light curve extraction}
We begin with the uncalibrated detector files [\texttt{uncal.fits}].
We initially process these using the stage 1 steps from the \texttt{jwst} pipeline\footnote{\url{https://jwst-pipeline.readthedocs.io/en/latest/}} (v1.13.4) and apply a custom $1/f$ correction prior to the \texttt{ramp\_fit} step, as in other applications of \tiberius\ \citep[e.g.,][]{kirkBOWIEALIGNJWSTComparative2024}.
For this, we first fit the centre of the aperture trace with a fourth order polynomial, and create a custom mask with a defined width centred on the fitted trace. 
In each column, we take the median of the remaining unmasked pixels in the detector image and subtract it from the column.
We then run the \texttt{ramp\_fit} and \texttt{gain\_scale} steps, combining the 80 groups per integration.

To extract the light curves, we begin by applying the \texttt{Tiberius} cosmic ray correction to the \texttt{gainscalestep.fits} products; for each pixel we take a running median over every three integrations and flag any integrations in which the pixel value is more than four standard deviations away from that median.
The flagged pixels are replaced by their median values.
We take the cosmic ray-corrected \texttt{gainscalestep.fits} 2D images, and fit the stellar aperture trace.
Using a user-defined initial guess row location and search width, first \texttt{Tiberius} fits each pixel column (in the cross-dispersion direction) with a Gaussian to find the peak flux.
Then, we fit a fourth order polynomial through the peaks of the fitted Gaussian for smoothing.
To extract the stellar flux, we replace any bad pixels in the 2D image (the bad, saturated, hot and dead pixels flagged in the \texttt{dq\_init} step of stage 1, combined with the \texttt{Tiberius}-identified $5\sigma$-clipped pixels) with the median of the pixels to the left and right.
We define an aperture width of 6 pixels (chosen to minimise the noise in the white light curve), and perform aperture photometry, summing the flux centred on the fitted aperture trace for each integration.
We extracted the stellar flux over pixels [600, 2040] and [5, 2040] inclusive (in the dispersion direction) for detectors NRS1 and NRS2 respectively.
The result is a time-series of stellar spectra.
For the wavelength solution, we make use of the \texttt{assign\_wcs} product from the stage 1 \texttt{jwst} processing.
The white light curves are shown in Fig.~\ref{fig:WLCs}, having integrated over wavelengths 2.75--3.72\,microns and 3.82--5.17\,microns for NRS1 and NRS2 respectively.

We create the spectroscopic light curves by integrating over a defined pixel bin width; we use the binning schemes from \citet{kirkBOWIEALIGNJWSTReveals2025}, creating spectra at resolutions of $R\simeq100$ (30/31, $\sim$60 pixel-wide wavelength bins for NRS1/2) and $R\simeq400$ (120/121, $\sim$15 pixel-wide bins).
In addition, we create `high-resolution', pixel-level light curves which will be analysed in a future publication.

{%
\renewcommand{\arraystretch}{1.4}
\begin{table*}

    \centering
    \caption{Best-fit planet parameters from the extracted white light curves. We only provide the weighted mean values for comparison; only the individual fit values highlighted in bold were used in this work.}
    \begin{threeparttable}

    \begin{tabular}{l c c c c c c c}
    
    \toprule
     &  & & $t_\mathrm{mid}$\,[BJD$_\mathrm{TDB}$] & $a/R_*$  & $i$\,[deg] & $R_\mathrm{p}/R_*$ \\
     
    \midrule
    Literature value & &~&  $60478.21264\pm0.00045$\,$^*$ & $6.04\pm0.23$\,$^*$ & $83.1^{+0.5}_{-0.4}$\,$^*$  & $0.10452^{+0.00066}_{-0.00072}\,^\vdag$ \\
    \midrule
    \tiberius\ & priors & --& $|t_\mathrm{mid}-60478.21|<0.1$ & $a/R_* \ge 1$ & \begin{tabular}{@{}c@{}}$i \sim \mathcal{U}[0,90]$ \\ $i-i_0>-5^\ddagger$\end{tabular}
    & $R_\mathrm{p}/R_* \sim \mathcal{U}[0,0.5]$ \\
    
    & \textbf{individual} & \textbf{NRS1}  & $\mathbf{60478.212976}{\substack{\mathbf{+0.000039}\\\mathbf{{-0.000040}}}}$ & $\mathbf{5.941 \pm 0.016}$& $\mathbf{82.574}{\substack{\mathbf{+0.035}\\\mathbf{-0.034}}}$& $\mathbf{0.100064}{\substack{\mathbf{+0.000069}\\\mathbf{-0.000067}}}$\\
    & \textbf{individual} & \textbf{NRS2}  & $\mathbf{60478.212839\pm0.000038}$& $\mathbf{5.945}{\substack{\mathbf{+0.017}\\\mathbf{-0.016}}}$ &$\mathbf{82.591 \pm0.035}$& $\mathbf{0.099260\pm0.000066}$\\
    & weighted mean & -- & $60478.212885\pm0.000023$ & $5.943\pm0.011$ & $82.582\pm0.025$ & $0.099650\pm0.000047$\\
    
    \midrule 
    \eureka\ & priors & -- & $t_\mathrm{mid} \sim \mathcal{N}[60478.21,0.1]$ & $a/R_* \sim \mathcal{N}[6.04,1.0]$ & $i \sim \mathcal{U}[80,90]$ & $R_\mathrm{p}/R_* \sim \mathcal{N}[\delta_0,0.1]^\ddagger$\\
    & individual & NRS1 & $60478.213010 \pm 0.000060$ & $5.925\pm 0.029$ & $82.528^{+0.067}_{-0.066}$ & $ 0.10005 \pm 0.00021$\\
    & individual & NRS2 & $60478.213100{\substack{+0.000077\\-0.000079}}$ &  $5.945^{+0.029}_{-0.028}$ & $82.59^{+0.066}_{-0.063}$ & $ 0.09952 \pm 0.00024$\\
    & weighted mean & -- & $60478.213043\pm0.000048$ & $5.935\pm0.020$ & $82.560\pm0.046$ & $0.09982\pm0.00016$\\

    \bottomrule
    \end{tabular}
    \begin{tablenotes}
        \item[*] Values from \citet{kokoriExoClockProjectIII2023}.
        The mid-transit time from \citet{kokoriExoClockProjectIII2023} has been propagated to the epoch of our JWST observation, accounting for the uncertainties on the literature mid-transit time and period (provided in Table~\ref{table:system_params});
        \item[$\vdag$] Values from \citet{sozzettiGAPSProgrammeHARPSN2015};
        \item[$\ddagger$] $i_0$ is the initialised value, for which we use the literature value listed above. $\delta_0$ is the literature value for $R_\mathrm{p}/R_*$. 
    \end{tablenotes}
    
    \end{threeparttable}
    \label{table:retrieved-params}

\end{table*}
}
\begin{figure}
    \centering
    \includegraphics[width=0.98\linewidth]{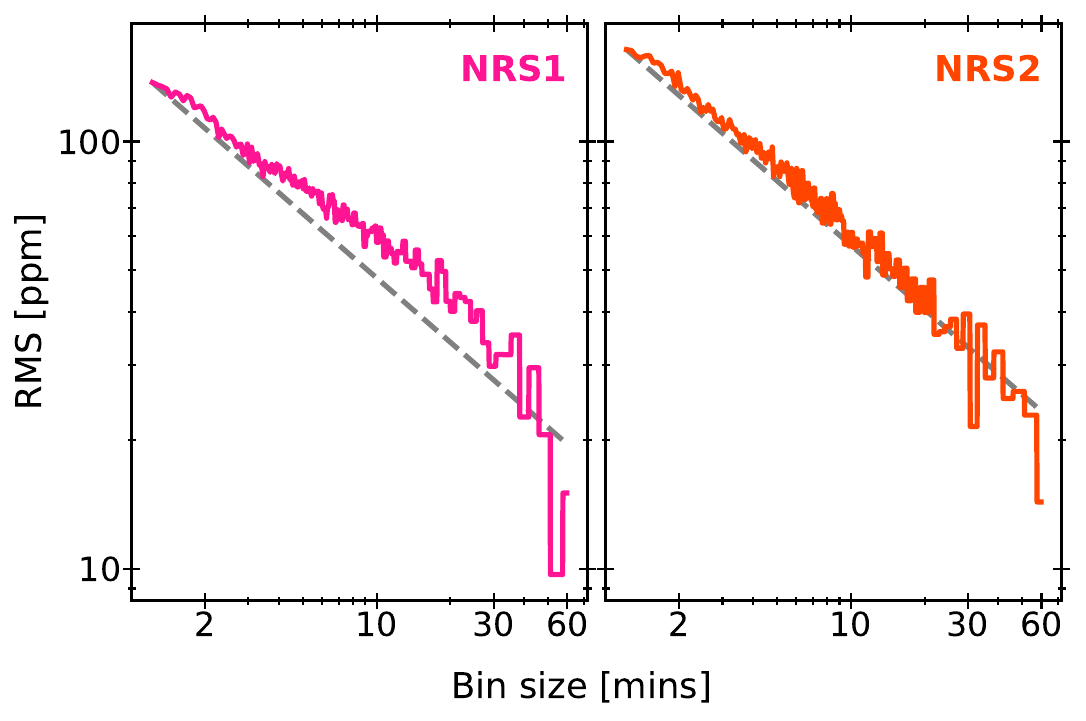}
    \caption{Allan variance plot for the individually fit \texttt{Tiberius} light curves presented in Fig.~\ref{fig:WLCs}. The grey dashed lines show the expectation from white noise.}
\label{fig:allan}
\end{figure}

\subsubsection{Light curve fitting}
\label{section:tiberius-LC-fitting}
For the fitting of the light curves, we zero-centre the time arrays on the predicted mid-transit time, propagated from the latest published ephemeris \citep[Table~\ref{table:system_params} and~\ref{table:retrieved-params};][]{kokoriExoClockProjectIII2023}.
Prior to fitting, we applied sigma clipping along the time axis with a $4\sigma$ threshold to remove outliers.
The light curve model consisted of a transit light curve model \citep[using the \texttt{batman} package;][]{kreidbergBatmanBAsicTransit2015} multiplied by a systematics model.
For the latter, we found a linear polynomial in time to sufficiently detrend the data, affording the lowest Bayesian Information Criterion (BIC) compared to more complex models tested.
This contributed two free parameters, the polynomial coefficients, $c_n$; we used wide, uniform priors, $|c_n|<10$.
For the transit model, we leave $t_\mathrm{mid}$, $a/R_*$, $i$ and $R_\mathrm{p}/R_*$ as free parameters, again applying only loose, wide priors (given in Table~\ref{table:retrieved-params}).
We fixed the orbital period to $P=3.55392889$\,days \citep[][]{kokoriExoClockProjectIII2023}.
We also fix the eccentricity and longitude of periastron to $e=0$ and $\omega=90^\circ$, as consistent with literature values (see Table~\ref{table:system_params}).
Assuming a quadratic limb-darkening law, we used \texttt{ExoTiC-LD} with the 3D \textsc{stagger} stellar atmosphere model grid \citep[][]{grantExoTiCLDThirtySeconds2024,magicStaggergridGrid3D2015} to generate limb-darkening coefficients for the published stellar parameters and fixed these in our light curve fitting (NRS1: $u_1=0.0658$, $u_2=0.1117$; NRS2: $u_1=0.0560$, $u_2=0.0928$).

We leveraged \texttt{emcee} \citep[][]{Foreman-Mackey2013EmceeHammer} to explore the posterior distribution, with 30 walkers per free parameter. 
We run an initial burn-in with $10,000$ steps which are discarded, rescale the photometric uncertainties to give a reduced chi-square $\chi^2_\nu=1$, then run $10,000$ production steps.
In Fig.~\ref{fig:WLCs} we show the \texttt{Tiberius}-extracted white light curves, and their best-fit models, having fit the detectors separately.
The associated Allan variance plots are shown in Fig.~\ref{fig:allan} \citep[][]{allanStatisticsAtomicFrequency1966}.
The parameters recovered from the transit light curve models are given in Table~\ref{table:retrieved-params}, and the posterior distributions are displayed in Fig.~\ref{fig-app:wlc-corner}.

\begin{figure}
    \centering
    \includegraphics[width=0.98\linewidth]{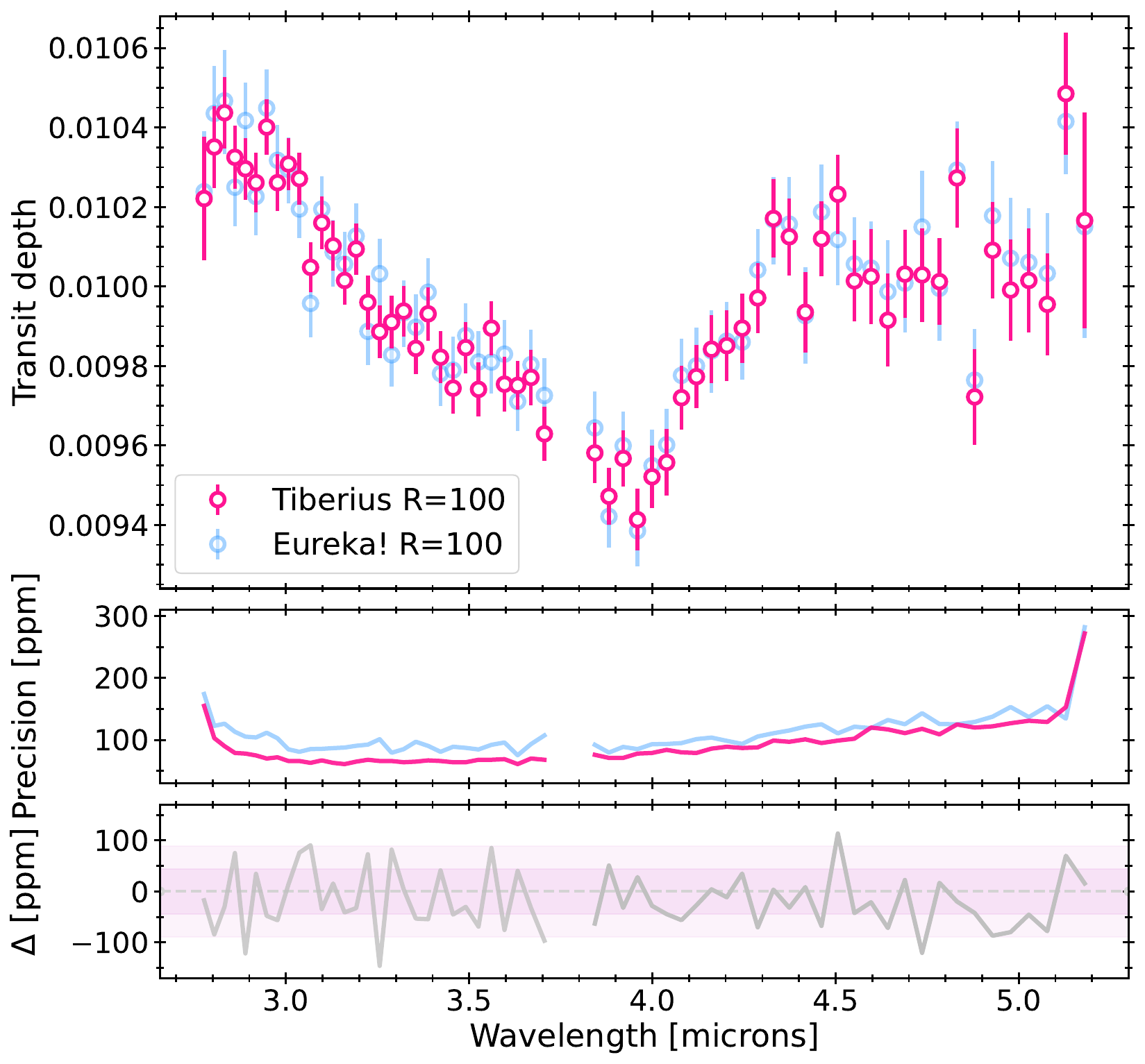}
    \caption{\textbf{Top panel:} The $R\simeq100$ transmission spectrum of \planet from the \texttt{Tiberius} and \texttt{Eureka!} reductions in pink and blue respectively. \textbf{Middle panel:} The precision of each spectrum and \textbf{bottom panel:} the difference between the two reductions. The $1/2\sigma$ intervals are shaded for reference.}
    \label{fig:transmission-spec}
\end{figure}

We proceed to fit the wavelength-binned light curves, with linear systematics models for all wavelength bins.
Again, we fixed the quadratic limb-darkening coefficients to that pre-computed with the same \textsc{stagger} grid, for the bin-centre wavelengths.
Having fixed the $t_\mathrm{mid}$, $a/R_*$ and $i$ to the fitted white light curve values (using the individual detector parameters; Table~\ref{table:retrieved-params}), we fit $R_\mathrm{p}/R_*$ for each light curve. 
For the spectroscopic light curve fits, we employ a Levenberg-Marquadt algorithm to optimise the model parameters \citep[][]{More1978}, scaling the uncertainties to give $\chi_\nu^2=1$ and re-running the fit as done for the white light curves.
The resulting \texttt{Tiberius} $R\simeq100$ transmission spectrum is shown in pink in Fig.~\ref{fig:transmission-spec}, with associated precision shown in the middle panel.
The $R\simeq400$ spectrum is shown in Fig.~\ref{fig-app:R400} of Appendix~\ref{SECTION-8:appendix-SLCs}.

To comment on the residuals from the white light curve model fits (right panel, Fig.~\ref{fig:WLCs}), we notice possible correlated noise at time $-1.1$\,hrs and $+0.4$\,hrs from mid-transit. 
These features are more pronounced in the NRS1 light curve.
Indeed, looking at the Allan variance in Fig.~\ref{fig:allan} there is some residual red noise in NRS1, while the residual noise closely resembles what we would expect of photon noise for NRS2.
Such wavelength dependence may support a stellar origin, namely occultations of photospheric features, rather than a systematic origin.
To investigate this, we attempted to add a spot component to our transit model, with 4 additional free parameters: the $(x,y)$ co-ordinates of the spot centre, the spot radius and the spot contrast.
We found no evidence for a spot-occultation in the white light curve.
Further, we saw no evidence of a wavelength-dependent feature, as would be expected from a spot-crossing event, in the residuals of the spectroscopic light curves.
We thus conclude these features are unlikely to be caused by occultations of star spots.

\subsection{\texttt{Eureka!}}
\subsubsection{Light curve extraction}
We started our \texttt{Eureka!} reduction with Stage\,1 and Stage\,2, which are wrapped around the default \texttt{jwst} pipeline\footnote{\url{https://jwst-pipeline.readthedocs.io/en/latest/}} (v1.12.2). 
Within this wrapper and before the \texttt{ramp\_fit} step, we applied a custom $1/f$ correction at the group level which uses the Stage\,3 column-by-column background subtraction setup.
We subtracted a zero-order polynomial that was fitted to the background area (having masked the trace) and rejected outliers $>$3$\times$ the median absolute deviation. 
We opted to increase the jump detection step at Stage\,1 to avoid false positives, as the default is often too low (set to 10.0 instead of 4.0) and we skipped the \texttt{photom\_step} at Stage\,2. 

Using \texttt{Eureka!}'s Stage\,3 we extracted the stellar spectra. 
Within this stage we corrected for the curvature of the trace by identifying the central trace pixel in each column and moving the spectra such that the central trace pixel aligns. 
We performed column-by-column background subtraction: using the pixels more than 6 pixels away from the centre of the trace, we employ an outlier rejection threshold of 5$\times$ the median in the spatial direction and a $7\sigma$ along the time axis following two iterations.
Finally, we used optimal spectral extraction \citep[][]{Horne1986AnSpectroscopy.} over a total aperture size of 9 pixels, to extract the time-series spectra over pixels [600, 2040] and [5, 2040] for detectors NRS1 and NRS2, respectively. 

We bin our spectra into light curves using \texttt{Eureka!}'s Stage\,4 at resolutions of $R\simeq100$ and $R\simeq400$, equal to the binning scheme used in the \texttt{Tiberius} reduction and in previous BOWIE-ALIGN analyses. 
We utilise a 5$\sigma$-clipping from a rolling median of 25 pixels to mask outliers in the light curves (0-1 outliers per light curve in both unbinned and binned cases). 
In addition, we adopted manual masking prior to Stage\,4 to mask bad wavelength columns since we found some outliers from Stage\,3 were not masked sufficiently, and were propagating through to the light curves and transmission spectrum. 
For that we identified outliers in the spectra, which differed more than $20\times$ the uncertainty from the rolling mean of 25\,pixels.  

\begin{figure*}
    \includegraphics[width=0.98\linewidth]{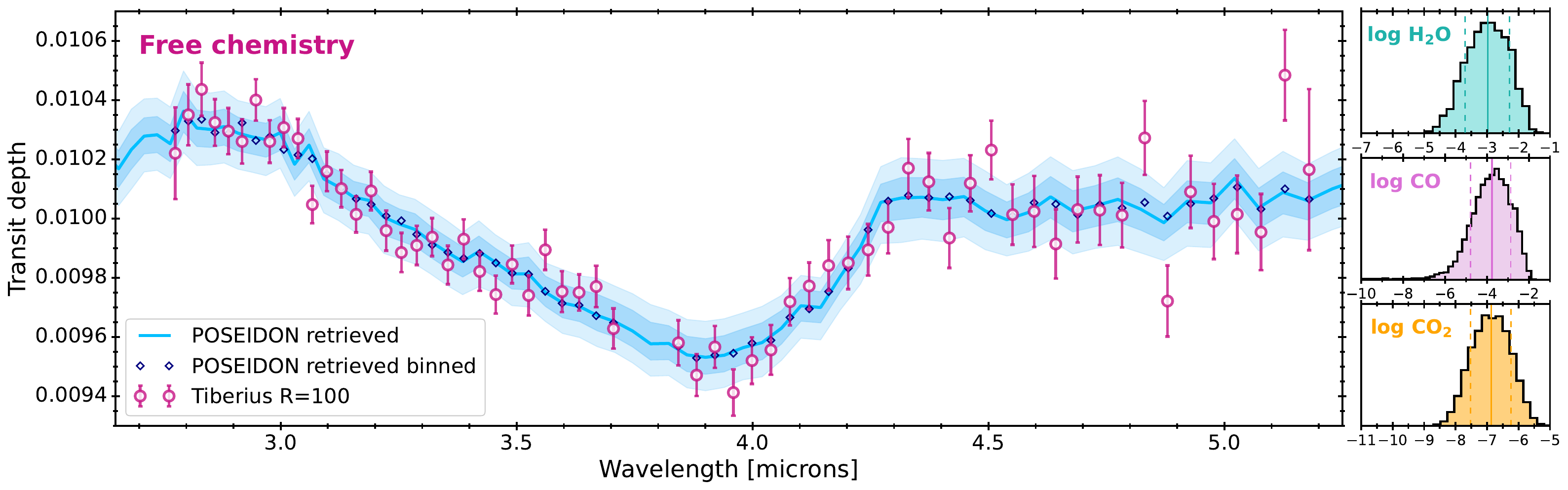}
    \caption{\textbf{Main:} Median retrieved \texttt{POSEIDON} spectrum for the free chemistry retrieval, smoothed to a resolution of $R=100$, in blue (with $1/2\sigma$ intervals shaded in successively lighter blue), for the \texttt{Tiberius} $R\simeq100$ transmission spectrum (pink data points). 
    The median retrieved spectrum convolved with the instrument point spread and sensitivity function, then binned to the data resolution is also shown in the navy diamond data points. 
    \textbf{Right:} Posterior distributions for the log volume mixing ratios of \ch{H2O} (top), \ch{CO} (middle) and \ch{CO2} (bottom).} 
    \label{fig:poseidon-free}
\end{figure*}

\subsubsection{Light curve fitting}

Using \texttt{Eureka!}'s Stage\,5 we fitted our light curves. 
First, we fit the white light curves of both NRS1 and NRS2 separately, freely fitting for the orbital parameters $t_\mathrm{mid}$, $a/R_*$, $i$ and $R_\mathrm{p}$/$R_*$ and using a simple linear-in-time systematics model. 
As with the \texttt{Tiberius} reduction, we fix the orbital period to $P = 3.55392889$\,days \citep[][]{kokoriExoClockProjectIII2023} and eccentricity to zero. 
For the limb-darkening, we used the quadratic limb-darkening law and left both $u_1$ and $u_2$ as free parameters, using a wide uniform prior.
For all light curve fitting within \texttt{Eureka!}, we utilise the \texttt{batman} package \citep[][]{kreidbergBatmanBAsicTransit2015} for our transit model and the Markov Chain Monte Carlo (MCMC) package \texttt{emcee} \citep{Foreman-Mackey2013EmceeHammer} to retrieve our fitted parameters. 
We used 50 walkers per free parameter, 20,000 steps for the white light curve fit (with 10,000 steps of that discarded as burn-in) and 1,000 steps for the spectroscopic light curves (with 500 steps of that discarded as burn-in). 
The fitted values from the white light curves using \texttt{Eureka!} can be found in Table\,\ref{table:retrieved-params}. 

For the purpose of comparing the transmission spectra directly between the \texttt{Tiberius} and \texttt{Eureka!} reductions we fixed the system parameters ($i$, $a/R_*$ and $t_\mathrm{mid}$) to the \texttt{Tiberius} values for the spectroscopic light curves. 
This leaves six free parameters for each spectroscopic light curve: $R_\mathrm{p}$/$R_*$, limb-darkening coefficients $u_1$ and $u_2$, the two linear trend coefficients and an error inflation term.
The resulting \eureka\ $R\simeq100$ transmission spectrum is shown in Fig.~\ref{fig:transmission-spec} alongside the equivalent \texttt{Tiberius} spectrum; the $R\simeq400$ is shown in Appendix~\ref{SECTION-8:appendix-SLCs}, Fig.~\ref{fig-app:R400}.

\subsection{The transmission spectrum of \planet}

We show the transmission spectrum of \planet\ from each pipeline, at a spectral resolution of $R\simeq100$, in the top panel of Figure~\ref{fig:transmission-spec}.
The $R\simeq400$ transmission spectra are shown in Appendix~\ref{SECTION-8:appendix-SLCs}, Fig.~\ref{fig-app:R400}.
In Appendix~\ref{SECTION-7:appendix-tiberiusJK} we detail a third independent reduction, a second implementation of \tiberius\ (v2).
The spectra from this reduction are also shown in Fig.~\ref{fig-app:R400}, and the $R\simeq100$ spectrum compared to the result from the primary \tiberius\ reduction in Fig.~\ref{fig-app:tiberius_comparison}.

We see excellent agreement between the independent reductions.
Indeed, the median differences between the $R\simeq100$ spectra were $6$ and $28$\,ppm for NRS1 and NRS2 (bottom panel Fig.~\ref{fig:transmission-spec}), which were lower than the median precisions ($79$ and $101$\,ppm for \tiberius\ and \eureka\ respectively; see middle panel Fig.~\ref{fig:transmission-spec}).

One key difference to highlight between the \tiberius\ and \eureka\ light curve-fitting routines was the treatment of limb darkening.
In the primary \tiberius\ reduction (\S\ref{section:tiberius-LC-fitting}), we used a quadratic limb-darkening law, with fixed limb-darkening coefficients.
We also used a quadratic limb-darkening law in the fitting of the \eureka\ light curves, but the coefficients were left as free parameters.
We obtained consistent transmission spectra, as described above, with larger uncertainties on the \eureka\ transit depths, which would be consistent with the higher model dimensionality.
We found similar inflation of the transit depth uncertainties in the secondary \tiberius\ reduction, described in Appendix~\ref{SECTION-7:appendix-tiberiusJK}, when fitting for the limb-darkening coefficients.
In summary, our tests attest to the robustness of the transmission spectrum against choice of limb-darkening treatment.
We proceed to analyse the primary \tiberius\ and the \eureka\ transmission spectra using atmospheric retrievals.

\section{Retrievals}

\label{SECTION-4:retrievals}
In order to interpret our transmission spectra, we perform a series of atmospheric retrieval tests on both the \tiberius\ and \eureka\ transmission spectra, with the publicly available packages \poseidon\ and \prt.
The configuration and priors used for each retrieval are summarised in Tables~\ref{table-app:retrieval-priors} and \ref{table-app:retrieval-priors-stellar-activity}.
We detail each in turn below.

{%
\renewcommand{\arraystretch}{1.2}
\begin{table*}
    \caption{Retrieval results: mean and $1\sigma$ confidence intervals of chemical mixing ratios, $\log X_i$, from the free chemistry retrievals, as presented in \S\ref{SECTION-4:retrievals}. For non-detected species, we provide the $2\sigma$ upper limits.
    We also provide values for the chemical equilibrium and hybrid chemistry retrievals, listing the mean and $1\sigma$ uncertainties, at the pressure of maximum contribution (see e.g., Fig.~\ref{fig-app:madhu-PT}).}
    \label{tab:all_retrievals}
    \begin{tabular}{l c c c c c c c } 
    \toprule
     Input spectrum  & H$_2$O & CO  & CO$_2$ & CH$_4$ & H$_2$S & SO$_2$ & HCN \\
     \midrule
     \poseidon  & & & & & & & \\
     \textit{Free Chemistry}  & & & & & & & \\

     \tiberius, $R=100$  & $-2.98 {\substack{+0.68\\-0.73}}$ & $-3.76{\substack{+0.89\\-1.01}}$ & $-6.86{\substack{+0.62\\-0.65}}$ & $<-7.07$ & $<-4.51$ & $<-6.97$ & $<-6.97$\\
     
     \tiberius, $R=400$  & $-2.88\pm0.61$ & $-3.97{\substack{+0.95\\-1.25}}$ & $-6.81{\substack{+0.56\\-0.54}}$&   $<-7.01$& $<-4.98$ & $<-7.25$ & $<-6.82 $\\

     \eureka, $R=100$ & $-2.88{\substack{+0.61\\-0.72}}$ & $-3.93{\substack{+0.89\\-1.03}}$ & $-7.11{\substack{+0.63\\-0.67}}$ &  $<-7.01 $&$<-5.18$ & $<-7.38$ & $<-7.06$\\

     \eureka, $R=400$ & $-2.89{\substack{+0.59\\-0.53}}$ & $-3.71{\substack{+0.89\\-1.03}}$ & $-6.94\pm0.56$& $<-7.47 $ & $<-5.41$ &$<-7.37$ & $<-6.20$\\

     \textit{Equilibrium Chemistry}  & & & & & & & \\
     \tiberius, $R=100$  & $ -3.40{\substack{+0.26\\-0.23}}$ & $-3.46{\substack{+0.24\\-0.26}}$ & $-7.04\pm0.43$ & $-9.38{\substack{+0.69\\-0.68}}$ & $-4.77{\substack{+0.23\\-0.20}}$ & $-13.31{\substack{+1.05\\-1.02}}$ & $-10.48{\substack{+0.26\\-0.25}}$\\
     
    \tiberius, $R=400$  & $-3.32{\substack{+0.27\\-0.24}}$ & $-3.56{\substack{+0.26\\-0.32}}$ & $-7.02{\substack{+0.51\\-0.49}}$ & $-9.20{\substack{+0.64\\-0.75}}$ & $-4.75{\substack{+0.24\\-0.25}}$ & $-13.48{\substack{+1.16\\-1.09}}$ & $-10.67{\substack{+0.24\\-0.22}}$\\
     
     \midrule 
     \prt & & & & &   \\

     \textit{Free Chemistry} & & & & & & & \\

     \tiberius, $R=100$ & $-3.29{\substack{+0.57\\-0.59}}$ & $-3.71{\substack{+0.85\\-0.84}}$ & $-7.21{\substack{+0.56\\-0.51}}$ &  $<-7.54 $ & $<-4.87$ & $<-7.24$ & $<-7.03$\\
     
     \tiberius, $R=400$ & $-3.26{\substack{+0.51\\-0.56}}$ & $-3.74{\substack{+0.80\\-0.83}}$ & $-7.13{\substack{+0.49\\-0.52}}$ &  $<-7.20$ & $<-4.57$ & $<-7.12$ & $<-7.14$\\

     \eureka, $R=100$ &  $-3.18{\substack{+0.57\\-0.65}}$ & $-3.31{\substack{+0.84\\-0.95}}$ & $-7.19{\substack{+0.59\\-0.64}}$ &  $<-7.53$ & $<-5.17$ & $<-7.03$ & $<-7.03$\\

     \textit{Hybrid Chemistry} &  & & & & & & \\

     \tiberius, $R=100$ &  $ -3.49{\substack{+0.24\\-0.25}}$ & $-3.74{\substack{+0.22\\-0.26}}$ & $-7.30{\substack{+0.48\\-0.47}}$ & $-11.78{\substack{+0.69\\-0.76}}$ & $<-4.70$ & $<-6.90$ & $-12.28{\substack{+0.26\\-0.23}}$\\
     
     \tiberius, $R=400$ &  $ -3.45{\substack{+0.23\\-0.25}}$ & $-3.75{\substack{+0.21\\-0.23}}$ & $-7.25{\substack{+0.45\\-0.48}}$ & $-11.68{\substack{+0.70\\-0.59}}$ & $<-4.64$ & $<-6.86$ & $-12.33{\substack{+0.25\\-0.22}}$\\

     \eureka, $R=100$ &  $ -3.61\pm0.26$ & $ -3.62{\substack{+0.27\\-0.25}}$ & $ -7.34{\substack{+0.51\\-0.46}}$ & $ -11.88{\substack{+0.80\\-0.61}}$ & $<-4.96$ & $<-6.74$ & $ -11.99{\substack{+0.30\\-0.27}}$\\

     \textit{Equilibrium Chemistry}  & & & & & & & \\
     \tiberius, $R=100$  & $ -3.48{\substack{+0.24\\-0.25}}$ & $-3.63{\substack{+0.22\\-0.26}}$ & $-7.20{\substack{+0.48\\-0.47}}$ & $-11.89{\pm0.69}$ & $-4.95{\substack{+0.21\\-0.20}}$ & -- & $-12.13{\substack{+0.28\\-0.27}}$\\
     
    \tiberius, $R=400$  & $-3.45{\substack{+0.29\\-0.25}}$ & $-3.67{\substack{+0.29\\-0.34}}$ & $-7.18{\substack{+0.50\\-0.48}}$ & $-11.75{\substack{+0.69\\-0.76}}$ & $-4.98{\substack{+0.23\\-0.25}}$ & -- & $-12.22{\substack{+0.26\\-0.23}}$\\
     
    \bottomrule
    \end{tabular}

\label{table:retrieval-results-species}
\end{table*}
}

\subsection{\texttt{POSEIDON}}
\label{section4:poseidon}
We leverage the \texttt{python} package \texttt{POSEIDON}\footnote{\url{https://poseidon-retrievals.readthedocs.io/en/latest/index.html}} \citep[v1.2.1;][]{macdonaldHD209458bNew2017,macdonaldPOSEIDONMultidimensionalAtmospheric2023} to perform free chemistry and equilibrium chemistry  atmospheric retrievals on the transmission spectrum of \planet.
\poseidon\ uses the forward-modelling tool \texttt{TRIDENT} \citep[][]{macdonaldTRIDENTRapid3D2022} to compute the transmission spectra, integrated with a nested sampling framework to explore the parameter space.
For the forward models, we use a resolution of $R=20,000$; high resolution is recommended when using opacity sampling \citep[see e.g.,][]{garlandEffectivelyCalculatingGaseous2019}.
We instantiate a \ch{H2/He}-dominated \citep[with fixed ratio \ch{He/H2} = 0.17;][]{asplundChemicalCompositionSun2009} atmosphere with trace gases \ch{CH4}, \ch{CO2}, \ch{H2O} \citep[ExoMol;][]{yurchenkoExoMolLineLists2024,yurchenkoExoMolLineLists2020a,polyanskyExoMolMolecularLine2018}, CO \citep[][]{liRovibrationalLineLists2015}, \ch{H2S} \citep[][]{azzamExoMolMolecularLine2016}, HCN \citep[][]{barberExoMolLineLists2014} and \ch{SO2} \citep[ExoAmes;][]{underwoodExoMolMolecularLine2016}.
We also adopt contributors to continuum absorption, namely collision-induced absorption (dominant for NIRSpec G395H wavelengths).
We use \ch{H2-H2}, \ch{H2-He}, \ch{H2-CH4}, \ch{CO2-H2}, \ch{CO2-CO2}, and \ch{CO2-CH4} cross-sections from HITRAN \citep[][]{karmanUpdateHITRANCollisioninduced2019,macdonaldTRIDENTRapid3D2022}.

The atmosphere of \planet\ is modelled with 100 layers evenly distributed in log-pressure, from $P_\mathrm{min}=10^{-7}$ to $ P_\mathrm{max}=10^2$\,bar.
In all of these retrievals, we allow the reference radius $R_\mathrm{p,ref}$ (defined at a reference pressure of $P_\mathrm{ref}=10$\,bar) to vary freely, having applied a uniform prior with a bounded range of 20\% the average of the fitted white light curve radius (Table~\ref{table:retrieved-params}).
Where specified, 
    \begin{enumerate}[label=(\roman*), leftmargin=2em, labelwidth=1.5em]
        \item we employ the \citet{macdonaldHD209458bNew2017} cloud prescription, selecting a cloud deck $+$ haze layer as defined by parameters: opaque cloud deck pressure, $P_\mathrm{cloud}$, the Rayleigh-enhancement factor, $a$, and a scattering slope, $\gamma$. 
        The haze opacity is parameterised as $\kappa_\mathrm{cloud} = a  \sigma_0 (\lambda/\lambda_0)^\gamma$ above the cloud deck pressure, $P_\mathrm{cloud}$ (below which the atmosphere is rendered opaque). The remaining parameters are constants: $\sigma_0$ is the \ch{H2} Rayleigh scattering cross section at reference wavelength $\lambda_0$ ($5.31\times 10^{-31}$\,m$^2$ and 350\,nm respectively).
        This cloud is uniform across the terminators. This adds three free parameters to the retrieval;
        \item we allow the $\log g$ of the planet to vary, having applied a Gaussian prior according to the solution given by the published mass \citep[$M_\mathrm{p}=0.494\pm0.035\,M_\mathrm{J}$;][]{sozzettiGAPSProgrammeHARPSN2015} and average of the radii measured from our \tiberius\ white light curves. This adds one free parameter to the retrieval;
        \item we include an offset between detectors NRS1 and NRS2, adding one free parameter.
    \end{enumerate}
Regarding the star, we fix all parameters to the literature values, provided in Table~\ref{table:system_params}.
We employ the nested sampling algorithm \texttt{PyMultiNest} \citep[][]{buchnerXraySpectralModelling2014,ferozMultiNestEfficientRobust2009}, with 1000 live points, to explore the parameter space.

\begin{figure}
\includegraphics[width=\columnwidth]{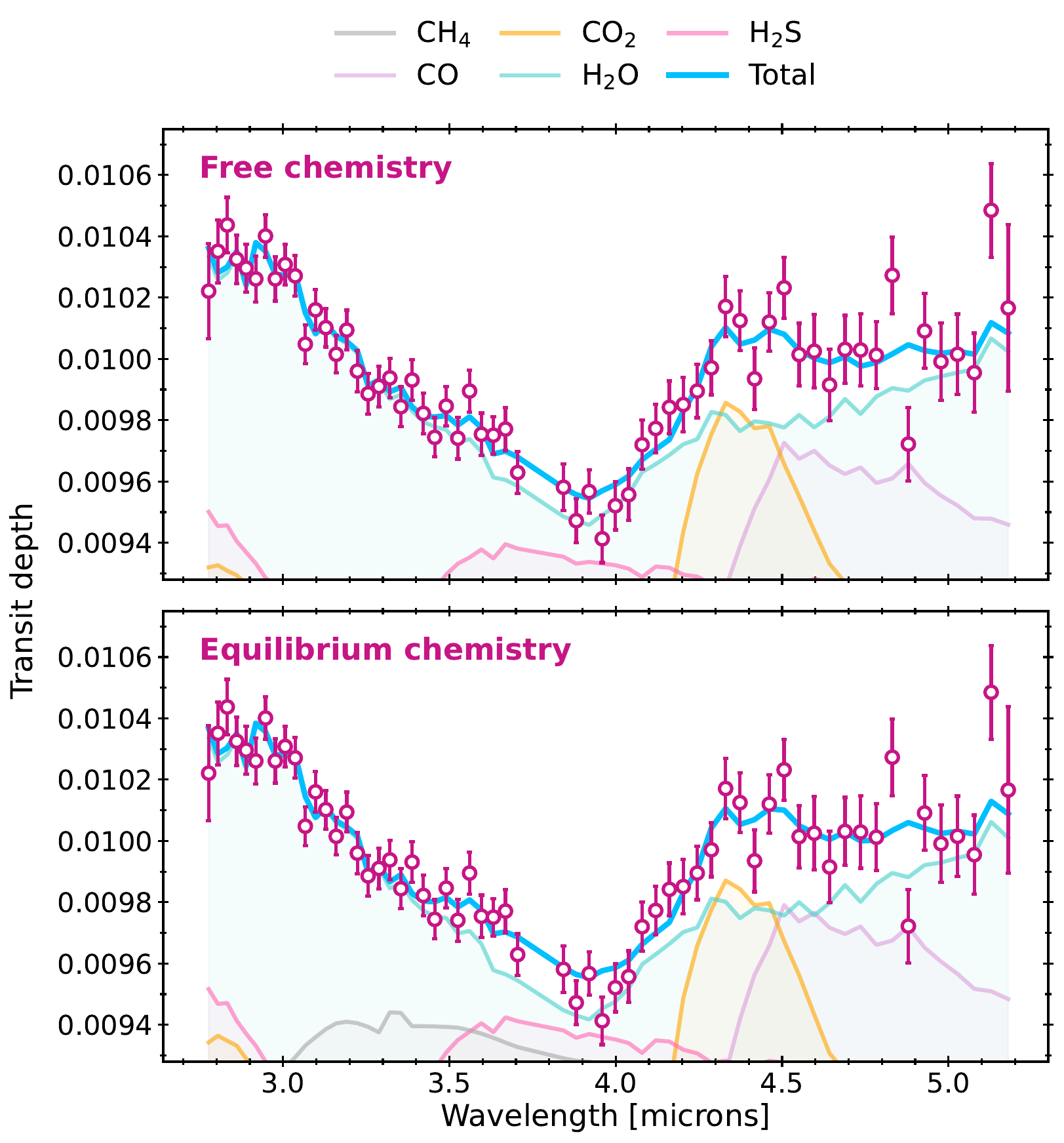}
\caption{Spectral contributions, convolved to the instrument PSF and binned to $R=100$, from the \textbf{(top)} free chemistry and \textbf{(bottom)} equilibrium chemistry \texttt{POSEIDON} retrievals on the $R\simeq100$ \texttt{Tiberius} spectrum.
The total spectrum is the median retrieved as shown in Fig.~\ref{fig:poseidon-free} and Fig.~\ref{fig:poseidon-chemeq-spec}. 
Though \ch{HCN} and \ch{SO2} were included in the equilibrium chemistry retrieval, their constrained contributions are not within the ranges plotted here.}
\label{fig:spec-contributions}
\end{figure}

\begin{figure*}
    \centering
    \includegraphics[width=0.98\linewidth]{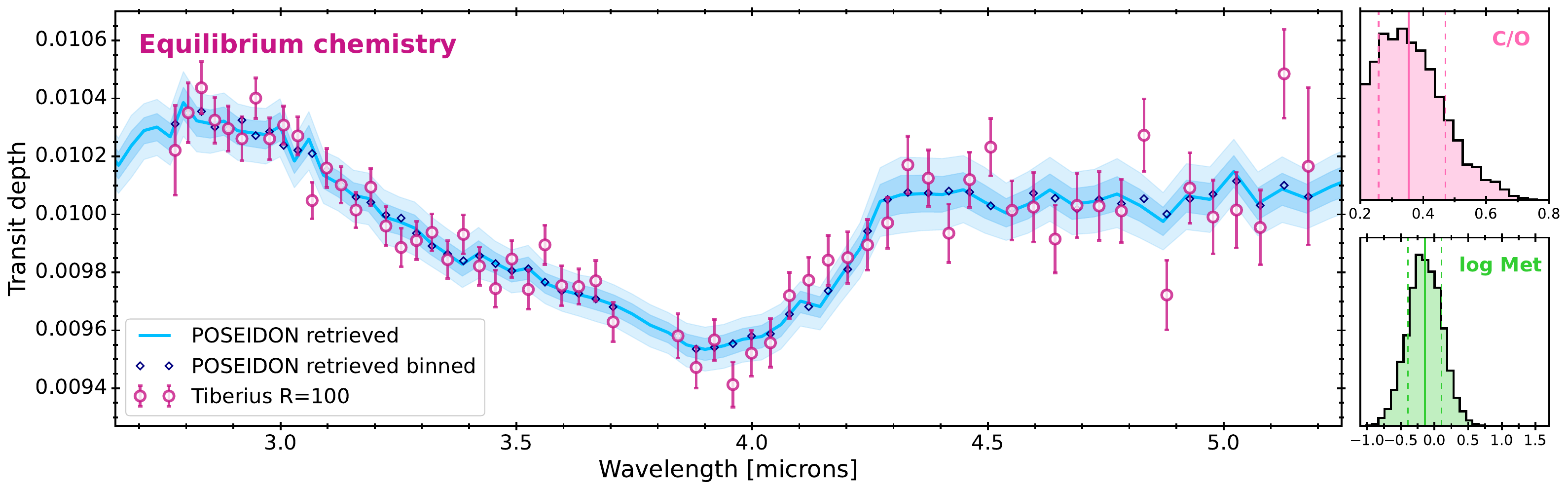}
    \caption{\textbf{Main:} Median retrieved equilibrium chemistry spectrum with \texttt{POSEIDON} in blue, in the same format as Fig.~\ref{fig:poseidon-free}, again for the \texttt{Tiberius} $R\simeq100$ transmission spectrum.
    \textbf{Right:} Posterior distributions for $\mathrm{C/O}$ (top), and metallicity (bottom). }
    
    \label{fig:poseidon-chemeq-spec}
\end{figure*}

\subsubsection{Free chemistry}
\label{section4:poseidon-free}
We first test atmospheric models wherein the abundances of the atmospheric species are allowed to vary freely.
For these free chemistry tests, we implement an isothermal pressure-temperature (PT) profile.
Our reference retrieval test:
\begin{enumerate}[label=(\Roman*),start=1,leftmargin=2em, labelwidth=1.5em]
    \item \label{free-baseline} a free chemistry retrieval including the species mentioned in the preceding section, a cloud deck and haze parameterisation, and an offset between detectors, totalling 14 free parameters.
\end{enumerate}
For each of the trace species, we assume constant abundance with pressure, and apply a wide uniform prior on the volume mixing ratio of each ($-12<\log_{10}X_i<-1$\,dex; see Table~\ref{table-app:retrieval-priors}).

\textbf{Test}~\ref{free-baseline}. 
On the $R\simeq100$ \tiberius\ spectrum, we retrieve abundances $\log X_\mathrm{H_2O} = -2.98{\substack{+0.68\\-0.73}}$, $\log X_\mathrm{CO}=-3.76{\substack{+0.89\\-1.01}}$ and $\log X_\mathrm{CO_2} = -6.86{\substack{+0.62\\-0.65}}$.
We plot the median retrieved spectrum in Fig.~\ref{fig:poseidon-free}, together with the posterior distributions for \ch{H2O}, \ch{CO} and \ch{CO2}.
We provide all reference retrieval results in Tables~\ref{table:retrieval-results-species} and \ref{table:retrieval-results} and the associated posterior distributions are shown in Fig.~\ref{fig-app:free-corner}, Appendix~\ref{SECTION-9:appendix-retrieval-results}.
The median retrieved spectrum from the reference retrieval \ref{free-baseline} yields a reduced chi-square of $\chi^2_\nu=1.27$, with 47 degrees of freedom.
In Fig.~\ref{fig:spec-contributions} we plot the contributions from individual gas opacities to the overall median retrieved spectrum.

To quantify the contribution of each parameter in the model, we run series of `nested model' retrievals, with one contribution removed at a time.
We assign detection significance by Bayesian model comparison.
Comparing the Bayesian evidence $\mathcal{Z}$ to that of the reference model, we compute a Bayes' factor which is then transformed to a frequentist detection `sigma' significance \citep[][]{trottaBayesSkyBayesian2008,bennekeHowDistinguishCloudy2013}.
This significance nominally quantifies the level at which the added complexity contributes to a better model fit, and is a standard metric in the reporting of exoplanet atmospheric characteristics \citep[see e.g.,][]{welbanksAuroraGeneralisedRetrieval2021,taylorAwesomeSOSSAtmospheric2023,bennekeHowDistinguishCloudy2013}.

For the \tiberius\ $R\simeq 100$ spectrum, we detect H$_2$O and CO$_2$ to $8.4\sigma$ and $4.0\sigma$ respectively ($12.3\sigma$ and $4.5\sigma$ for the $R\simeq 400$ transmission spectrum).
We detect CO to $3.1\sigma$ ($2.4\sigma$) but find insufficient evidence in favour of  CH$_4$, HCN, or SO$_2$.
We see possible hints of \ch{H2S}; Bayesian model comparison yielded a $1.2\sigma$ model preference. 
This is insufficient to claim a detection, and the posterior was non-Gaussian (Fig.~\ref{fig-app:free-corner}).
Still, we place a $2\sigma$ upper limit $\log X_{\rm{H_2S}} < -4.51$, i.e. $31$\,ppm; for reference, the solar sulphur abundance is approximately 13\,ppm \citep{asplundChemicalCompositionSun2009}.
We also investigated the presence of clouds in the atmosphere of \planet. 
In the reference retrieval \ref{free-baseline} on the \tiberius\ $R\simeq100$ transmission spectrum, an opaque cloud deck level was retrieved at a pressure of $\log P_\mathrm{cloud} = 0.20{\substack{+1.14\\-1.13}}$; looking at the pressure contribution function \citep[see Fig.~\ref{fig-app:madhu-PT};][]{mullensImplementationAerosolMie2024}, this cloud deck is below the pressures probed and not likely to provide a significant contribution to the observed spectrum.
Indeed, the evidence for the baseline model is $\ln \mathcal{Z} = 468.87 \pm 0.12$, and then $\ln \mathcal{Z} = 469.96\pm 0.12$ when we remove the cloud $+$ haze layer from the model.
There is therefore insufficient evidence to support the presence of a cloud deck or haze layer, and we place a conservative $2\sigma$ upper limit of $\log P_\mathrm{cloud}\,\rm{[bar]} > -1.70$.
Further, the reference retrieval recovers an offset between detectors NRS1 and NRS2 that is consistent with zero ($\delta_\mathrm{rel}=-10^{+466}_{-476}$\,ppm), though not well constrained.
We note that when removing the offset parameter from the retrieval, we recover chemical abundances that are completely in agreement with the reference retrieval.
We therefore do not see evidence for a detector offset in the transmission spectrum.

We run the same retrieval tests on both of the \eureka\ transmission spectra.
The baseline model \ref{free-baseline} from the reference retrieval results in a reduced chi-square of $\chi^2_\nu=1.45$, with Bayesian evidence $\ln \mathcal{Z} = 452.54\pm0.12$.
We retrieved H$_2$O, CO and CO$_2$ at abundances comparable to those retrieved for the \tiberius\ spectra, at significances of $9.8\sigma$, $2.9\sigma$ and $3.0\sigma$ respectively for the $R\simeq100$ \eureka\ spectrum ($9.3\sigma$, $2.8\sigma$ and $3.6\sigma$ for the \eureka\ $R\simeq400$ spectrum).
In this case, we also find no evidence for a detector offset, clouds, CH$_4$, H$_2$S, HCN, or SO$_2$.

\subsubsection{Chemical equilibrium}
\label{section4:poseidon-chemeqm}
We independently test atmospheric compositions wherein the relative abundances of the atmospheric species are governed by  equilibrium chemistry.
\poseidon\ interpolates over a pre-computed chemistry grid from \texttt{FastChem} \citep{stockFastChemComputerProgram2018,stockFastChem2Improved2022} for the equilibrium calculations.
We employ the subset of chemical species indicated in \S\ref{SECTION-4:retrievals}, and a gradient PT profile, linearly varying with $\log P$ between two temperatures, $T_\mathrm{deep}$ and $T_\mathrm{high}$.
We apply equal priors to the high and deep temperatures, with no user-defined prior on $\Delta T=T_\mathrm{deep}-T_\mathrm{high}$ or ${\delta T}/ {\delta P}$ (Table~\ref{table-app:retrieval-priors}).
Our reference retrieval includes:
\begin{enumerate}[label=(\Roman*),start=2,leftmargin=2em, labelwidth=1.5em]
    \item \label{eqm-baseline} an atmosphere under equilibrium chemistry, with a cloud deck and haze layer. We include an offset between detectors. Since the chemistry is entirely defined by the carbon-to-oxygen ($\mathrm{C/O}$) ratio and the metallicity relative to solar metallicity ($Z/Z_\odot$)\footnote{Throughout this work, the $Z/Z_\odot$ notation refers to metallicity set by C/H, i.e., $[\mathrm{M/H}]_\mathrm{C/H}$.}, we have a total of 10 free parameters.
\end{enumerate}
We used uniform priors for $\mathrm{C/O}$ and metallicity, across the full ranges supported by \poseidon\ ($\mathrm{C/O}\sim\mathcal{U}[0.2, 2.0]$; $[\mathrm{M/H}]_\mathrm{O/H} \sim \mathcal{U}[-1, 4]$).

\begin{figure}
    \centering
    \includegraphics[width=0.98\linewidth]{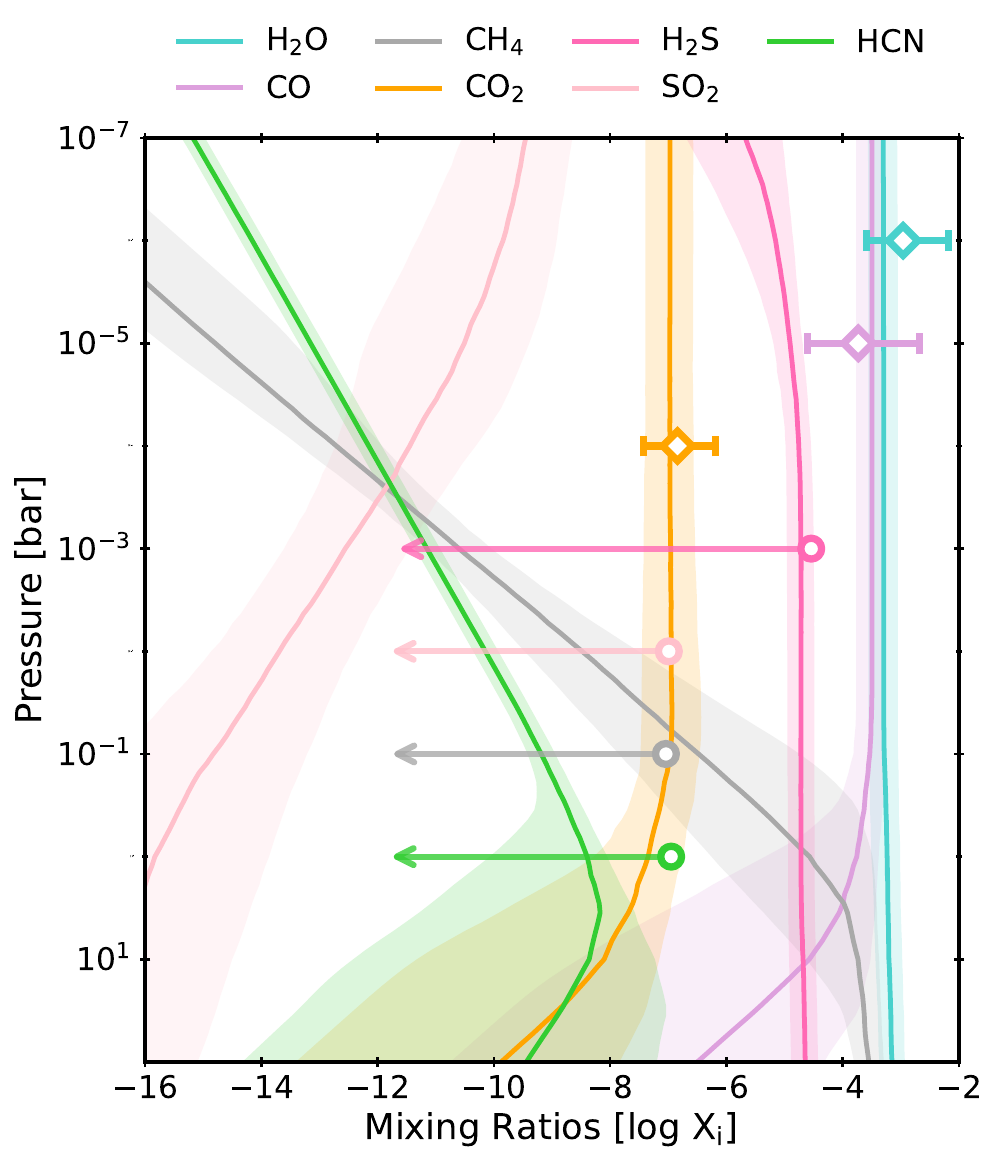}
    \caption{Vertical chemical profiles (median and $1\sigma$ confidence intervals) of the trace species in chemical equilibrium, from the best-fit \texttt{POSEIDON} retrieval on the $R\simeq100$ \texttt{Tiberius} transmission spectrum (see \S\ref{section4:poseidon-chemeqm}). We retrieve a $\mathrm{C/O}=0.35^{+0.12}_{-0.10}$ and $\log Z/Z_\odot =-0.15{\substack{+0.27\\-0.26}} $. Abundance constraints from the free chemistry retrieval on the same transmission spectrum in \S\ref{section4:poseidon-free} are overplotted (at arbitrary pressures): we retrieve bounded constraints on H$_2$O, CO and CO$_2$ (diamond markers).
    We display the $2\sigma$ upper limits on the other, non-detected species (circle markers).}
    \label{fig:poseidon_chemeqm_profiles}
\end{figure}

\textbf{Test}~\ref{eqm-baseline}. 
For the $R\simeq100$ \tiberius\ transmission spectrum, we retrieve $\mathrm{C/O}=0.35^{+0.12}_{-0.10}$ and $\log Z/Z_\odot=-0.15{\substack{+0.27\\-0.26}}$ ($\mathrm{C/O}=0.32^{+0.11}_{-0.08}$ and $\log Z/Z_\odot=-0.23{\substack{+0.30\\-0.28}}$ for the $R\simeq 400$ spectrum)\footnote{We note that we have converted the retrieved \poseidon\ metallicities to the same scale as \prt, where the metallicity is set by C/H, and the O/H is then calculated from the metallicity and the C/O ratio, listed as $[\mathrm{M/H}]_\mathrm{C/H}$, for ease of comparison. 
The metallicity posterior in Fig.~\ref{fig-app:chemeq-corner} is a direct output of \poseidon, with the metallicity set by O/H.
We discuss this further in \S\ref{section5:retrieval-diff}.}.
We retrieve a reference radius and planetary $\log g$ comparable to those of the free retrieval.
The median retrieved spectrum is shown in Fig.~\ref{fig:poseidon-chemeq-spec}, along with posterior distributions for $\mathrm{C/O}$, and metallicity (see Appendix~\ref{SECTION-9:appendix-retrieval-results}, Fig.~\ref{fig-app:chemeq-corner} for the full corner plot), and yields a reduced chi-square of $\chi_\nu^2=1.13$ with 51 degrees of freedom.
In Fig.~\ref{fig:poseidon_chemeqm_profiles}, we show the vertical mixing profiles (mean and $1\sigma$ intervals) for the model in Fig.~\ref{fig:poseidon-chemeq-spec}, and provide the abundances at the maximum pressure contribution (see Fig.~\ref{fig-app:madhu-PT}) in Table~\ref{table:retrieval-results-species}.
The low abundances of (e.g.,) \ch{SO2} and HCN are not actually observed (see Fig.~\ref{fig:spec-contributions}), but rather constrained by the equilibrium model abundances.

We note that we did investigate a more complex PT profile, namely the \citet{Madhusudhan2009} profile.
This parameterisation divides the atmospheres into three layers, provisionally allowing for a temperature inversion within the middle layer, and has 6 free parameters.
We show the median retrieved PT profile along with the transmission spectrum contribution function, and the simpler gradient profile in Fig.~\ref{fig-app:madhu-PT}.
We saw no evidence for a temperature inversion, and notably the retrieved limb temperature is in agreement with that retrieved using the simpler gradient PT profile ($T_\mathrm{ref}=1313{\substack{+104\\-89}}$\,K at  $P_\mathrm{ref}=10$\,bar for the \citet{Madhusudhan2009} PT profile).
We retrieved completely consistent atmospheric parameters, including $\mathrm{C/O}=0.35^{+0.12}_{-0.09}$ and $\log Z/Z_\odot = -0.17{\substack{+0.27\\-0.25}}$.
We thus favour the model with the less complex gradient PT profile presented in the preceding paragraph, as recommended by \citet{schleichKnobsDialsRetrieving2024}.

We run the same reference retrieval tests on the \eureka\ spectra.
The reference retrieval on the $R\simeq100$ spectrum yields a $\mathrm{C/O}$ comparable to that retrieved for the equivalent \tiberius\ spectrum, $\mathrm{C/O}=0.40{\substack{+0.15\\-0.12}}$ ($\mathrm{C/O}=0.40{\substack{+0.13\\-0.11}}$ for the $R\simeq400$ spectrum), with the corresponding model $\chi^2_\nu=1.33$.
However, we recover a slightly varied metallicities for the \eureka\ spectra: $\log Z/Z_\odot = -0.26{\substack{+0.37\\-0.33}}$ and $\log Z/Z_\odot = -0.04\pm0.36$ for the $R\simeq 100$ and $400$ spectra respectively.
These are still consistent with those retrieved for the 
{%
\onecolumn
\begin{landscape}
\renewcommand{\arraystretch}{1.2}
\begin{table}

    \caption{Retrieval results (mean and $1\sigma$ confidence intervals) as presented in \S\ref{SECTION-4:retrievals}. For the unconstrained parameters, we provide the $2\sigma$ upper limits.}
    \begin{threeparttable}
    \label{table:retrieval-results}
    \begin{tabular}{l c c c c cc c c c c c c} 
    \toprule
     Input spectrum & \# & $\ln \mathcal{Z}$ & $R_\mathrm{p,ref}\,[R_\mathrm{J}]$ & $\log g$\,[cgs] &  \multicolumn{2}{c}{$T$\,[K]$^\vdag$}  & $\log P_\mathrm{cloud}$\,[bar]  &  $\log a$ & $\gamma$ & $\mathrm{C/O}$ & $\log Z/Z_\odot~^\ddagger$  & Offset\,[ppm] \\
     \midrule
     \poseidon & & & & & & & & & & &\\
     \textit{Free Chemistry} & & & & & & & & & & &\\
     
     \tiberius, $R=100$ & \ref{free-baseline} & $468.9\pm 0.1$ & $1.63{\substack{0.01\\-0.02}}$ & $2.57\pm0.04$ & 
      $1180{\substack{+139\\-138}}$ & --& 
     $>-1.70$ & $<7.41$ & $<-0.97$& -- & --& $<939$ \\

     \tiberius, $R=400$ &  & $1727.7\pm0.1$ & $1.64\pm0.01$ & $2.57{\substack{+0.04\\-0.03}}$ & $1089^{+132}_{-94}$ & -- &
     $>-1.44$ & $<7.56$ & $<-1.05$ & -- & -- & $<939$ \\

     \eureka, $R=100$ & & $452.5\pm0.1$& $1.65\pm0.01$ & $2.55\pm0.05$ & $1005{\substack{+193\\-144}}$ & -- & $>-1.32$ & $<7.43$ & $<-1.22$ & -- & -- & $<907$\\

     \eureka, $R=400$ & & $1710.3\pm0.1$ & $1.65 \pm0.01$ & $2.57{\substack{+0.04\\ -0.05}}$ & $981{\substack{+135\\ -110}}$ & -- & $>-1.17$ & $<7.43$ & $<-1.26$ & --& -- & $<934$\\

     \textit{Equilibrium Chemistry} & & & & & & & & &\\
     
     \tiberius, $R=100$ & \ref{eqm-baseline} & $473.3\pm 0.1$ & $1.61^{+0.03}_{-0.04}$ & $2.57{\substack{0.04\\-0.05}}$ & $1352{\substack{+149\\-184}}$ & $1235{\substack{+420\\-350}}$ & $0.21{\substack{+1.14\\-1.22}}$ & $1.08{\substack{+4.14\\-3.35}}$ & $-11.6{\substack{+6.0\\-5.4}}$ & $0.35{\substack{+0.12\\-0.10}}$ & $-0.15{\substack{+0.27\\-0.26}}$& $-6{\substack{+472\\-459}}$\\

     \tiberius, $R=400$ & & $1730.2\pm0.1$ & $1.61^{+0.02}_{-0.04}$ & $2.57^{+0.05}_{-0.06}$ & $1283^{+153}_{-254}$ & $1287^{+412}_{-358}$ & $0.23^{+1.18}_{-1.16}$ & $1.16^{+4.23}_{-3.42}$ & $-11.6^{+6.1}_{-5.4}$ & $0.32^{+0.11}_{-0.08}$ & $-0.23{\substack{+0.30\\-0.28}}$ & $-27^{+457}_{-452}$\\

     \eureka, $R=100$ & & $455.7\pm0.1$ & $1.61\pm0.03$ & $2.56{\substack{+0.06\\-0.07}}$ & $1279{\substack{+241\\-295}}$ & $1298{\substack{+388\\-364}}$ & $0.23{\substack{+1.16\\-1.13}}$ & $1.05{\substack{+4.13\\-3.32}}$ &$-11.7{\substack{+6.1\\-5.2}}$ & $ 0.40{\substack{+0.15\\-0.12}}$ & $-0.26{\substack{+0.37\\-0.33}}$ & $+20{\substack{+461\\-465}}$\\

     \eureka, $R=400$ & & $1714.0\pm 0.1$ & $1.62{\substack{+0.02\\-0.03}}$ & $2.62{\substack{+0.05\\-0.07}}$ & $1365{\substack{+195\\-313}}$ & $1329{\substack{+375\\-385}}$ & $0.22{\substack{+1.16\\-1.19}}$& $1.08{\substack{+4.12\\-3.38}}$& $-11.3{\substack{+5.8\\-5.4}}$& $0.40{\substack{+0.13\\-0.11}}$& $-0.04\pm0.36$ & $+13{\substack{+478\\-464}}$\\

     \tiberius, $R=100$ & \ref{eqm-contamination} & $473.2\pm 0.1$ & $1.61\pm0.03$ & $2.56{\substack{+0.04\\-0.05}}$ & $1324{\substack{+168\\-205}}$ & $1301{\substack{+385\\ -359}}$ & $0.14{\substack{+1.14\\-1.15}}$& $1.06{\substack{+3.77\\-3.20}}$ & $-11.9{\substack{+5.8\\-4.9}}$ & $0.37{\substack{+0.13\\-0.10}}$ & $-0.14{\substack{+0.27\\-0.26}}$ & $+30{\substack{+417\\-430}}$\\

     \midrule 
     \prt & & & & & &   \\
     \textit{Free Chemistry} & & & & & & & & & & &\\
    
     \tiberius, $R=100$ & \hyperlink{prt-free}{(IV)} & $-51.4\pm 0.1$ & $1.80\pm0.02$ & $2.55^{+0.03}_{-0.04}$ & 
     $1157^{+94}_{-123}$ & --& 
     $0.36^{+1.08}_{-1.21} $ & -- & -- & -- & -- & -- \\

     \tiberius, $R=400$ &  & $-143.9\pm0.1$ & $1.80\pm0.02$ & $2.54\pm0.03$ & $1132^{+88}_{-101}$ & -- &
     $0.28^{+1.09}_{-1.16}$ & -- & -- & -- & -- & -- \\

     \eureka, $R=100$ &  & $-49.3\pm 0.1$ & $1.80\pm0.02$ & $2.56\pm-0.04$ & 
     $1138^{+128}_{-133}$ & --& 
     $0.28^{+1.12}_{-1.17} $ & -- & -- & -- & -- & -- \\
     
     \textit{Equilibrium Chemistry} & & & & & & & & &\\

      \tiberius, $R=100$ & \hyperlink{prt-chemeq}{(V)} & $-44.5\pm0.1$ & $1.80\pm0.01$ & $2.52\pm0.03$ & $1223^{+116}_{-65}$ & -- & $0.29^{+1.18}_{-1.29}$ & -- & -- & $0.36^{+0.16}_{-0.12}$ & $-0.29^{+0.30}_{-0.23}$ & --\\

      \tiberius, $R=400$ & & $-139.1\pm0.1$ & $1.80\pm0.01$ & $2.52\pm0.03$ & $1198^{+78}_{-67}$ & -- & $0.31^{+1.14}_{-1.23}$ & -- & -- & $0.33^{+0.13}_{-0.11}$ & $-0.33^{+0.24}_{-0.19}$ & --\\

      \eureka, $R=100$ &  & $-44.9\pm 0.1$ & $1.80\pm0.02$ & $2.54\pm0.04$ & 
      $1241^{153}_{-81}$ & --& 
      $0.33^{+1.09}_{-1.27} $ & -- & -- & $0.42^{+0.20}_{-0.15} $ & $-0.28^{+0.35}_{-0.27} $ & -- \\

      \textit{Hybrid Chemistry} & & & & & & & & &\\
     
     \tiberius, $R=100$ & \hyperlink{prt-hybrid}{(VI)} & $-45.7\pm0.1$ & $1.80\pm0.01$ & $2.55\pm0.03$ & $1198^{+70}_{-63}$ & -- & $0.24^{+1.17}_{-1.22}$ & -- & -- & $0.32^{+0.13}_{-0.10}$ & $-0.40^{+0.24}_{-0.20}$ & --\\

     \tiberius, $R=400$ & & $-139.1\pm0.1$ & $1.80\pm0.01$ & $2.54^{+0.03}_{-0.04}$ & $1180^{+47}_{-59}$ & -- & $0.33^{+1.16}_{-1.25}$ & -- & -- & $0.30^{+0.11}_{-0.09}$ & $-0.41\pm0.18$ & --\\

     \eureka, $R=100$ &  & $-46.4\pm0.1$ & $1.81\pm0.01$ & $2.56\pm0.03$ & $1203^{+90}_{-67}$ & -- & $0.28^{+1.15}_{-1.19}$ & -- & -- & $0.38^{+0.14}_{-0.13}$ & $-0.39\pm0.25$ & --\\
      
    \bottomrule
    \end{tabular}
\begin{tablenotes}
    \item[$\vdag$] If one value given, an isothermal temperature profile was used. If two values given, a gradient PT profile was used, and the values are ordered as $T_\mathrm{high}$ and $T_\mathrm{deep}$;
    \item[$\ddagger$] The metallicities reported are set by $[\mathrm{C/H}]$ (native for \prt\ and converted for \poseidon\ ).
    \item \ref{free-baseline} and \ref{eqm-baseline} are the \poseidon\ reference free and chemical equilibrium retrievals presented in \S\ref{section4:poseidon-free} and \S\ref{section4:poseidon-chemeqm} respectively;
    \item \ref{eqm-contamination} is the \poseidon\ stellar contamination retrieval presented in \S\ref{section4:poseidon-chemeqm-activity};
    \item \hyperlink{prt-free}{(IV)}, \hyperlink{prt-chemeq}{(V)}, and \hyperlink{prt-hybrid}{(VI)} are the \prt\ free, equilibrium, and hybrid chemistry retrievals respectively, presented in \S\ref{section:prt-retrievals}.

\end{tablenotes}
\end{threeparttable}

\end{table}
\end{landscape}
\twocolumn
}

\noindent \tiberius\ spectra within $1\sigma$.
Again, all the retrieved parameters are given in Table~\ref{table:retrieval-results}.

\subsubsection{Stellar heterogeneities}
\label{section4:poseidon-chemeqm-activity}
Stellar surface heterogeneities have the potential to affect the observed transmission spectrum of a transiting planet.
Unocculted dark spots and bright regions introduce a wavelength-dependent flux variation, and since the transit depth is inferred assuming the planet atmosphere is entirely illuminated by the homogeneous photospheric spectrum, this can leave artefacts in the transmission spectrum.
This is a known phenomenon, dubbed the `transit light source effect' \citep[TLSE;][]{rackhamTransitLightSource2018,rackhamTransitLightSource2019}.
This has been shown to introduce degeneracies and even entirely account for transmission spectra in the near-infared, including with NIRSpec G395H observations \citepalias[see e.g.,][]{moranHighTideRiptide2023}.
Though one might not expect significant contributions from heterogeneities on the F-type host, we nevertheless run a test retrieval with a contamination factor included, considering the residual variability seen in the white light curves (\S\ref{section:tiberius-LCs}).
To test for stellar contamination:
\begin{enumerate}[label=(\Roman*),start=3,leftmargin=2em, labelwidth=1.5em]
    \item \label{eqm-contamination} we model one photospheric heterogeneity feature with an assumed \textsc{phoenix} model stellar spectrum defined by a characteristic temperature $T_\mathrm{het}$, and gravitational field strength $\log g_\mathrm{het}$. 
    The planet atmosphere is assumed to be under equilibrium chemistry and is parameterised as described in \S\ref{section4:poseidon-chemeqm}.
\end{enumerate}
Effectively it is a heterogeneity-photosphere temperature contrast over which we marginalise, since we also fit for the equivalent photospheric characteristics, $T_\mathrm{phot}$ and $\log g_\mathrm{phot}$. 
Therefore, along with the heterogeneity filling factor $f_\mathrm{het}$, this introduces five additional free parameters to the retrieval.
Using \poseidon\''s in-built functionality, we compute the transmission spectrum contamination factor defined in \citet{rackhamTransitLightSource2019}.
We run a retrieval, assuming chemical equilibrium, on the $R\simeq100$ \tiberius\ spectrum.
We apply heterogeneity priors $T_\mathrm{het} \sim \mathcal{U}[0.8T_*, 1.2T_*]$, and $\log g_\mathrm{het}\sim\mathcal{U}[3.0, 5.0]$ (see Table~\ref{table-app:retrieval-priors-stellar-activity} for the full list of priors), where we use the literature value for the stellar effective temperature, $T_*$ (Table~\ref{table:system_params}).
The definitions and priors of the other parameters explained above remain unchanged compared to the reference retrieval without stellar contamination (\S\ref{section4:poseidon-chemeqm}).
We thus have 15 free parameters.

We retrieve a possible bright feature at $T_\mathrm{het}= 6389{\substack{+388\\-313}}$\,K, with a filling factor of $f_\mathrm{het}=0.22{\substack{+0.21\\-0.13}}$.
The heterogeneity temperature is consistent with the retrieved photospheric temperature ($T_\mathrm{phot}= 6306{\substack{+56\\-60}}$\,K).

Moreover, the retrieved filling factor is relatively large considering the host stellar type; \citet{rackhamTransitLightSource2019} derive $f_\mathrm{fac}=0.01{\substack{+0.02\\-0.01}}$ for F6V stars.
It is also close to zero at the $1\sigma$ lower limit.
That said, there are other studies that have retrieved similarly high filling factors for other F stars (see \S\ref{section5:discussion-stellar-contam} for further discussion).
This retrieval yields a reduced chi-square of $\chi^2_\nu=1.21$, and the Bayesian evidence was $\ln \mathcal{Z}=473.16\pm0.11$.
The reference retrieval \ref{eqm-baseline} ($\ln \mathcal{Z}=473.35 \pm0.11$) without stellar contamination was thus marginally preferred to $1.4\sigma$.
We find that though a combination of a planetary atmosphere as well as stellar heterogeneities could explain the transmission spectrum of \planet, the reference model \ref{eqm-baseline} without stellar contamination is preferred due to the reduced model complexity.
We note that even when including the contamination from photospheric heterogeneities, the atmospheric parameters do not change; the $\mathrm{C/O}$ and metallicity are consistent with those from the chemical equilibrium reference retrieval \ref{eqm-baseline} (see Table~\ref{table:retrieval-results}).

\subsection{\texttt{petitRADTRANS}}
\label{section:prt-retrievals}

We also perform atmospheric retrievals on the \planet transmission spectrum using the \prt\footnote{\url{https://petitradtrans.readthedocs.io/en/latest/}} \citep[v2.7.7;][]{mollierePetitRADTRANSPythonRadiative2019,nasedkinAtmosphericRetrievalsPetitRADTRANS2024} package, with a near identical setup to that used to analyse the spectrum of the first BOWIE-ALIGN target, WASP-15\,b \citep[][]{kirkBOWIEALIGNJWSTReveals2025}. 
We use free chemistry, equilibrium chemistry, and hybrid chemistry retrieval setups assuming a \ch{H2}/He-dominated atmosphere, with absorption from trace gases computed from $R=1000$ correlated-$k$ opacity tables of the following atmospheric species: \ch{CH4} \citep{yurchenkoHybridLineList2017}, \ch{H2O} and CO \citep{Rothman2010}, \ch{CO2} \citep{yurchenkoExoMolLineLists2020a}, \ch{H2S} \citep{azzamExoMolMolecularLine2016}, \ch{SO2} \citep{underwoodExoMolMolecularLine2016}, and HCN \citep{barberExoMolLineLists2014}, plus collision-induced absorption from \ch{H2-H2} and \ch{H2-He}, and Rayleigh scattering from \ch{H2} and He. 

For each retrieval we model the atmosphere using 100 equal-log-spaced pressure layers from $10^{-6}\,$bar to 10$^2$\,bar and assume an isothermal pressure-temperature profile, with a wide uniform prior on the isothermal temperature from 500--3000 K. 
We use a wide uniform prior for the planetary reference radius (0.8--1.8 $R_\mathrm{J}$) and a Gaussian prior for gravity based on the published mass and radius \citep{sozzettiGAPSProgrammeHARPSN2015}, both of which are defined at a fixed reference pressure of 10$^{-3}$\,bar. 
We determined the appropriate reference pressure by running a retrieval with a Gaussian prior on the radius, based on the mean transit depth across both detectors, and the reference pressure as a free parameter. 
We also include an opaque grey cloud deck, with a log-uniform prior on the cloud-top pressure, from $10^{-6}\,$bar to 10$^2$ bar.

We run our retrievals on the \tiberius\  $R\simeq100$ and $R\simeq400$ spectra, as well as the \eureka\ spectrum at $R\simeq100$. 
Across these three retrieval tests, we retrieve consistent posteriors for radius, gravity, isotherm temperatures, and cloud-top pressure. 
We find a much lower limb temperature of $T=1200\pm100$\,K compared to the planet's equilibrium temperature of $T_\mathrm{eq}=1800$\,K.
This is a commonly observed feature in 1D isothermal retrievals \citep[see e.g.,][]{kirkBOWIEALIGNJWSTReveals2025,welbanksMassMetallicityTrendsTransiting2019}, which is postulated to arise from model choices such as the PT profile parametrization \citep{welbanks2022atmospheric} or unresolved limb asymmetries \citep{macdonald2020so}. 
The former was ruled out in \S\ref{section4:poseidon-chemeqm}, wherein we tested a more complex PT profile.
We also retrieve a relatively cloud-free atmosphere in all cases, placing a 3$\sigma$ upper limit on the grey cloud-top pressure of 10$^{-3}$\,bar.

\hypertarget{prt-free}{(IV)}
For the free chemistry retrievals, the abundance of each species is a free parameter, with a wide uniform prior in log mass fraction from $-12$ to $-0.5$. 
The free chemistry retrievals between all reductions and resolutions were consistent (see Table~\ref{table:retrieval-results-species}), revealing absorption due to \ch{H2O}, \ch{CO}, and \ch{CO2}, with retrieved abundances of $\log X_\mathrm{H_2O} = -3.29{\substack{+0.57\\-0.59}}$, $\log X_\mathrm{CO}=-3.71{\substack{+0.85\\-0.84}}$ and $\log X_\mathrm{CO_2} = -7.21{\substack{+0.56\\-0.51}}$.
We also see hints of \ch{H2S} abundances of 10\,ppm, but with a less than 2$\sigma$ confidence interval. 
We place 3$\sigma$ upper limits on the \ch{SO2} and HCN abundances of 4\,ppm and 2\,ppm respectively. The posterior distributions are presented in Appendix~\ref{SECTION-9:appendix-retrieval-results} Fig.~\ref{fig-app:pRT-free-corner}.

\hypertarget{prt-chemeq}{(V)}
For the equilibrium chemistry retrievals, the abundances of \ch{CH4}, \ch{H2O}, CO, \ch{CO2}, HCN, and \ch{H2S} are calculated at each pressure layer by interpolating a pre-computed equilibrium chemistry table with temperature, C/O ratio, and metallicity ([M/H]). We then use a wide uniform prior in both C/O ratio ($\mathrm{C/O}\sim\mathcal{U}[0.1, 1.5]$) and $\mathrm{[M/H]}\sim\mathcal{U}[-2, 3]$). 
In \prt, [M/H] is defined in terms of the solar values from \citet{asplundChemicalCompositionSun2009}, fixing the C/H value, with the O/H value then further modified by adjusting the C/O ratio (solar $\mathrm{C/O}\sim0.55$). 
We retrieve a sub-solar C/O ratio posterior of $0.33^{+0.13}_{-0.11}$ and a sub-solar metallicity of $-0.33^{+0.24}_{-0.19}$ for the \tiberius\ $R\simeq400$ spectrum, and consistent values at $R\simeq100$ with slightly reduced precision (see Table~\ref{table:retrieval-results}). 
We retrieve a similar but slightly higher C/O ratio of $0.43^{+0.17}_{-0.14}$ from the \eureka\ reduction.
We show the resulting posterior distributions from these retrievals in Appendix~\ref{SECTION-9:appendix-retrieval-results}, Fig.~\ref{fig-app:pRT-chemeq-corner}.

\hypertarget{prt-hybrid}{(VI)}
For the hybrid chemistry retrievals, we use the same equilibrium chemistry method to compute the abundances of \ch{CO2}, CO, \ch{H2O}, \ch{CH4}, and HCN, but allow the sulfur species \ch{H2S} and \ch{SO2} to vary freely, with broad uniform priors.
This is to allow the abundances to be in disequilibrium due to photochemical production of \ch{SO2} and destruction of \ch{H2S}, and to permit measurement of sulphur enrichment or depletion. 
As with our free chemistry retrieval, we see no evidence for \ch{SO2}, and hints of \ch{H2S}. 
As seen with WASP-15\,b \citep[][]{kirkBOWIEALIGNJWSTReveals2025}, moving to a hybrid retrieval results in tighter constraints on the C/O ratio; we retrieve a posterior of $0.30^{+0.11}_{-0.09}$. 
We also retrieve a somewhat lower metallicity of $-0.41\pm0.18$. 
The hybrid retrieval posteriors are presented in Appendix~\ref{SECTION-9:appendix-retrieval-results}, Fig.~\ref{fig-app:pRT-hybrid-corner}.
The Bayesian evidence for the equilibrium and hybrid retrievals are identical for the $R\simeq400$ retrievals, giving us no cause to favour one parametrization over the other, likely due to the absence of \ch{SO2} in the spectrum and the poor constraints on \ch{H2S}. At $R\simeq100$, the hybrid retrieval is slighly disfavoured, this difference is likely due the increased sensitivity to \ch{H2S} at higher resolutions.

\subsection{Comparison of retrieval results}
\label{section5:retrieval-diff}
As standard in the community, we used two independent retrieval analyses to assess the atmospheric composition of \planet.
This enables a robustness check of our inferences.
For ease of comparison, we plot the retrieved spectra from \poseidon\ and \prt\ together in Fig.~\ref{fig-app:all-retrieved-spec}.
We saw good agreement between the retrieved chemical abundances from the free chemistry retrievals (Table~\ref{table:retrieval-results-species}).
At face value, the chemical equilibrium retrievals of \poseidon\ and \prt\ yielded slightly discrepant atmospheric metallicity outputs.
This is due to the difference in metallicity prescriptions.
In \poseidon\ the abundances of all species except carbon are set by the metallicity, which is scaled from the solar value from \citet{asplundChemicalCompositionSun2009}.
The $[\mathrm{C/H}]$ is then set according to the $[\mathrm{O/H}]$ and $\mathrm{C/O}$ ratio.
Conversely, \prt\ sets all but the oxygen abundance based on the metallicity, and the $[\mathrm{O/H}]$ is derived.
To enable a direct comparison of the retrieved metallicities from both frameworks, we converted the \poseidon\ output $[\mathrm{M/H}]_\mathrm{O/H}$ to $[\mathrm{M/H}]_\mathrm{C/H}$.
Such a scaling is deemed reasonable for this work, since we only detect carbon and oxygen-bearing species, and the main observables (\ch{CO}, \ch{CO2} and \ch{H2O}) are not sensitive to the abundances of the other species.

Our \prt\ retrievals all used isothermal temperature profiles, previously thought to be sufficient for transmission spectroscopy.
Recently, \citet{schleichKnobsDialsRetrieving2024} showed that an isothermal prescription can lead to incorrect inferences and may be insufficient in the era of JWST exoplanet spectra.
With our \poseidon\ chemical equilibrium retrievals, we followed their recommendation of a two-point gradient temperature profile (sufficient except in the case of a temperature inversion, which we ruled out in \S\ref{section4:poseidon-chemeqm}).
However, we found only a small difference between the upper and deeper temperatures, maximally $347$\,K which is comparable to the $1\sigma$ constraints on $T_\mathrm{deep}$.

Ultimately, all retrievals yielded metallicities (i.e., $[\mathrm{M/H}]_\mathrm{C/H}$) consistent within $1\sigma$.
To summarise our collection of retrieval results, the metallicity of \planet\ is consistent with being sub-solar, and certainly lower than the metallicity of the host star ($\mathrm{[Fe/H]}=0.28$).
The retrieved sub-solar C/O was also consistent across all retrievals, varying between $\mathrm{C/O} = 0.30-0.42$.

\section{Discussion}

\label{SECTION-5:discussion}

\begin{figure}
 \centering
    \includegraphics[width=0.98\linewidth]{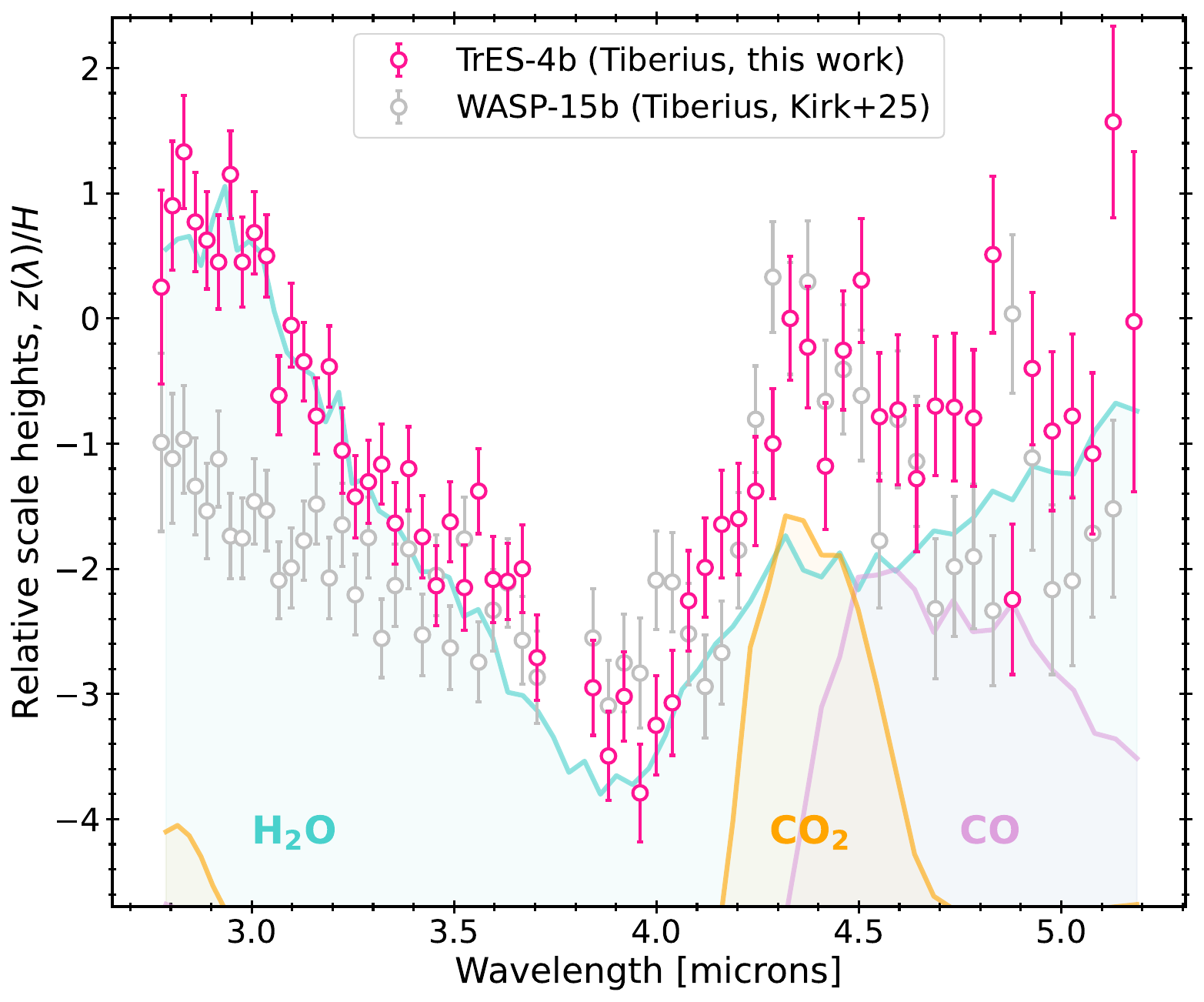}
    \caption{
    The transmission spectrum of \planet\ from this work (pink, \tiberius, $R\simeq100$) compared to the equivalent for WASP-15\,b, published in \citet{kirkBOWIEALIGNJWSTReveals2025} (grey).
    The spectral contributions from the detected gas opacities are overplotted for \planet.}
    \label{fig:vs_wasp15b}
\end{figure}

\subsection{The atmosphere of TrES-4 b}
We have analysed the atmosphere of \planet\ with JWST NIRSpec G395H data, two independent data reduction pipelines (\S\ref{SECTION-3:data-reduction}), together with two retrieval packages (\S\ref{SECTION-4:retrievals}).
Both retrieval frameworks yield consistent results, and our tests in \S\ref{SECTION-4:retrievals} present clear evidence for gaseous H$_2$O and CO$_2$ in the atmosphere of \planet.
Additionally, there is strong evidence for the presence of CO.
Considering the Bayesian evidence, the favoured \poseidon\ atmospheric model is that inferred by the chemical equilibrium retrieval \ref{eqm-baseline}.
In the \prt\ retrievals, the chemical equilibrium atmosphere was preferred by the $R\simeq100$ spectrum, while at $R\simeq 400$ the equilibrium and hybrid retrievals yielded equal evidence.
Neither \poseidon\ nor \prt\ found evidence of clouds or hazes.
We also tested for sulphur-bearing molecules \ch{H2S} and \ch{SO2}, finding weak evidence of \ch{H2S}, a non-detection of \ch{SO2}.
This is consistent with the lower retrieved metallicity \citep[see e.g.,][]{tsaiPhotochemicallyProducedSO22023,polmanPhotochemicallyProducedSO22023}.

\planet is the second planet from the BOWIE-ALIGN sample; the first, WASP-15\,b, was published in \citet{kirkBOWIEALIGNJWSTReveals2025}.
WASP-15\,b is a misaligned hot Jupiter, and was shown to host a super-solar metallicity atmosphere with $\mathrm{C/O}$ consistent with the solar value.
In Fig.~\ref{fig:vs_wasp15b} we show the NIRSpec G395H \tiberius\ $R\simeq100$ transmission spectrum of WASP-15\,b from \citet{kirkBOWIEALIGNJWSTReveals2025}, against the equivalent spectrum of \planet\ from this work (\S\ref{section:tiberius-LC-fitting}, shown in Fig.~\ref{fig:transmission-spec}).
For ease of comparison, we have scaled the transit depth by that for one atmospheric scale height ($139$\,ppm and $200$\,ppm for WASP-15\,b and \planet respectively), and then normalised to the $4.33$\,micron bin (the peak of the \ch{CO2} bandhead).
The comparatively reduced metallicity for \planet\ is apparent, considering the amplitude of the \ch{CO2} feature relative to the blueward 
\ch{H2O} slope.
Further, \citet{kirkBOWIEALIGNJWSTReveals2025} presented evidence of sulphur chemistry, through absorption features at $4.0$ and $4.9$\,microns.
The \ch{SO2}-induced feature seen at $4.0$\,microns for WASP-15\,b is clearly absent in our target \planet. 
Our retrievals yielded little evidence towards sulphur-bearing species; allowing the abundances of \ch{SO2} and \ch{H2S} to vary freely in the \prt\ hybrid retrievals (to allow for depletion or enrichment) did not result in increased model evidence.
Finally, the peak at $4.9$\,microns seen in WASP-15\,b's transmission spectrum is apparently inverted in that of \planet.
This inverted feature is consistent across reductions, but has a weaker structure at higher resolution  (Fig.~\ref{fig-app:R400}).
\citet{kirkBOWIEALIGNJWSTReveals2025} explored the possibility of OCS as the contributing absorber for the feature in WASP-15\,b.
In the case of \planet, the two outlier points just shy of $4.9$\,microns are still consistent with the best-fit chemical equilibrium model within $3\sigma$ (Fig.~\ref{fig:poseidon-chemeq-spec}).
Given the lack of other sulphur species, we choose not to investigate the presence of OCS further for \planet.

\begin{figure*}
\includegraphics[width=0.7\linewidth]{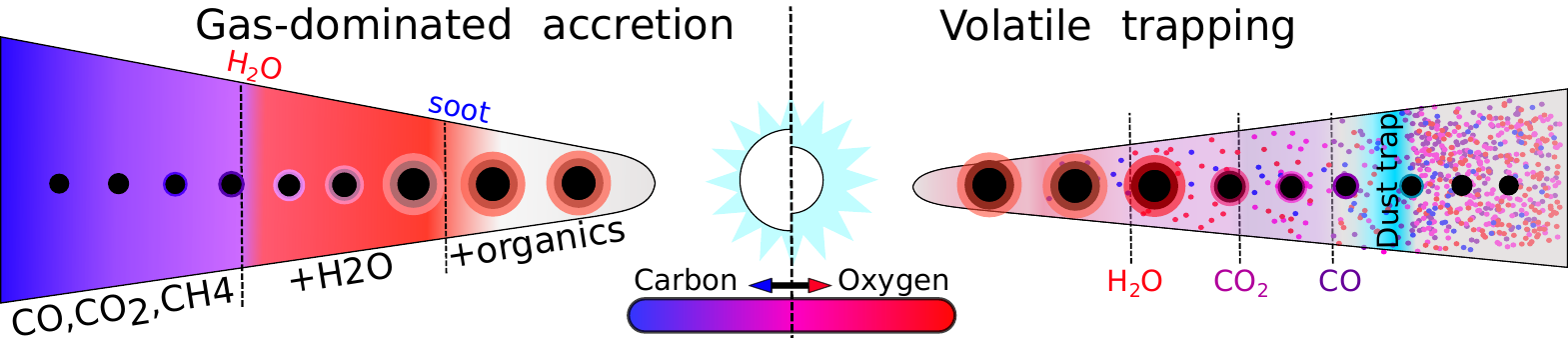}
\caption{Schematic of two distinct protoplanetary disc environment scenarios, in which \planet may have formed, relative to the host star in the centre. \textbf{Left:} The gas-dominated accretion scenario whereby the planet forms between the water ice line and the soot line (labelled dashed lines).
\textbf{Right:} The volatile trapping scenario, whereby the solids are trapped in the outer disc, rendering the inner disc lower in metallicity.
The relative abundances of carbon and oxygen are expressed by the colorbar.}
\label{fig:tres4-disc}
\end{figure*}

\subsection{Implications for formation scenarios}
Inferring formation histories for a single planet such as \planet is challenging, as results are typically degenerate and generally dependent on unknown factors such as the chemical composition of the disc (see, e.g., \citetalias{penzlinBOWIEALIGNHowFormation2024}).
Nevertheless, the sub-stellar metallicity and sub-solar C/O ratio place \planet in an interesting region of parameter space, which we consider here. 
In the typical picture of giant planet formation, gas accretion of the carbon-rich gases leads to low metallicities but high C/O, while strong solid accretion increases the metallicity through adding oxygen-rich ices and refractories, leading to low C/O and high metallicity.
In the \planet\ retrievals in \S\ref{SECTION-4:retrievals}, we retrieve a planet atmosphere metallicity between $\log Z/Z_\odot=-0.41$ to $-0.04$, which differs from the higher stellar metallicity at the $0.9\sigma$ to $3.4\sigma$ level.
Meanwhile, we find a C/O ratio between $0.30-0.42$ across all independent analyses. 
Considering the host star metallicity, a stellar C/O ratio in the range 0.4--0.5 is likely, based on observations of similar stars \citep[][]{Brewer2016}.
This would be compatible with the observed C/O ratio of \planet\ at $0.5-1.5\sigma$. 
We therefore cannot completely rule out the possibility that the planet's composition is close to the stellar composition, in which case no special circumstances are needed to explain it.

Conversely, a sub-stellar metallicity combined with sub-stellar C/O is challenging to explain within the standard picture of planet formation, assuming typical disc compositions. 
A particular challenge is that the gas composition in discs is usually thought to have super-stellar C/O ratios but sub-stellar metallicities due to the preferential condensation of oxygen-rich ices \citep[e.g.][]{obergEffectsSnowlinesPlanetary2011,Bosman2021,Bergin2024}. 
As a result, planets with sub-stellar metallicities are generally expected to have super-stellar C/O ratios and vice versa \citep[e.g.][]{Madhusudhan2014}, because adding more solids decreases the C/O at the cost of higher metallicity.
This makes the preferred composition of \planet\ somewhat unexpected. 
To explain the preferred composition, we must either invoke conditions in the disc where the gas has sub-stellar C/O and metallicity (reversing the trend) or invoke transport processes.

If the planet did form in a location where the gas composition has low C/O, then this points to one of two possible scenarios. 
First, the planet could have accreted the majority of its observable atmosphere inside the water ice line but outside the `soot line', inside which carbon-rich grains are destroyed \citep[][]{berginExoplanetVolatileCarbon2023,chachanRadialGradientsDusttogas2021}.
We depict this scenario on the left side of Fig.~\ref{fig:tres4-disc}.
In this region, the solids can have high C/O ratios if carbon grains are sufficiently abundant. 
Here, most oxygen-rich species apart from silicates are in the gas phase, resulting in lower C/O ratios in the gas. 
Observations of the interstellar medium suggest that roughly half of the interstellar carbon is contained in refractory grains \citep{2015Mishra}, which is sufficient to explain the composition of \planet. 
This explanation would point to the planet forming in the inner disc, where temperatures are $200-500$\,K.
Alternatively, chemical kinetics models of protoplanetary discs show that after $\sim 5$\,Myr the gas phase C/O ratio can become low in the inner $\sim 1-10\,{\rm AU}$ region of a disc, as CO and CO$_2$ are destroyed \citep{2018Eistrup}. 
Hence, the planet could have formed further out, but later in the stellar lifetime. 
A slight wrinkle in the second scenario is that the models producing low C/O ratio gas in the disc only include chemical kinetics, and accretion/pebble migration may overwhelm chemical processes that require such long time-scales to operate \citep[e.g.][]{Booth2019}.

Alternatively, the stalling of solid drift could help explain the low metallicity and C/O ratio (right side, Fig.~\ref{fig:tres4-disc}). 
If the disc has a pebble trap in the outer system where all volatile species are frozen out in grains (including \ch{H2O}, \ch{CO2}, \ch{CO} and \ch{CH4}), the disc inside the trap is metal depleted. 
Under such conditions, the planet would need to accrete fewer solids to acquire the observed C/O ratio and thus could have a lower metallicity. 
This mechanism is sensitive to the location and onset of the trapping feature in the disk \citep{2024Mah}.
Enhancing transport processes would not help explain the proposed composition.
For example, although efficient pebble migration can produce discs with low C/O ratios in the inner region, they do this by depositing large amounts of water, resulting in high metallicities \citep{Booth2017,2021Schneider}.
Similarly, although planet migration can modify the planet's composition, the effects are not large enough to change the broad conclusion that planets with low metallicity would ordinarily be expected to have high C/O ratios \citepalias{Madhusudhan2014,penzlinBOWIEALIGNHowFormation2024}.
For all these processes the amount of carbon in solid `soot' is crucial.
Through depleting carbon from molecules that can sublimate at very low temperatures (e.g., CO) to long-chain complex molecules that require temperatures of several hundred Kelvin to sublimate, the distribution of carbon in solids within the disc can shift, affecting the C/O ratio of the close-in final planet. 
If carbon was predominately in solids, late accretion would significantly increase C/O. 
If C/O was predominately in gases and volatiles, gas dominated accretion lead to higher C/O ratios and vice versa.
With all the planets in the BOWIE-ALIGN programme \citep{kirkBOWIEALIGNJWSTComparative2024}, we will have a sample to investigate the carbon evolution and distribution during planet formation.

A strong constraint on the atmospheric sulphur abundance would help distinguish these scenarios because most sulphur is likely in refractory form in both protoplanetary discs and the solar system \citep{kamaAbundantRefractorySulfur2019,turriniTracingFormationHistory2021,chachanRadialGradientsDusttogas2021,2023Crossfield}. 
However, our retrievals (\S\ref{SECTION-4:retrievals})  find no traces of sulphur species, except for hints of \ch{H2S}, for which the upper limit is too high to provide a meaningful constraint on the formation history.

Although the formation history of neither \planet\ nor WASP-15\,b is certain, it is clear that there should be differences in their histories.
WASP-15\,b has a super-solar metallicity, a C/O to ratio close to solar and a presumed high sulphur abundance. 
In contrast, \planet has a sub-stellar metallicity and C/O ratio.
Whether or not these differences are due to their aligned vs misaligned natures remains to be seen, a point that may be elucidated once the full BOWIE-ALIGN sample \citep{kirkBOWIEALIGNJWSTComparative2024} has been analysed.

\subsection{Stellar contamination}
\label{section5:discussion-stellar-contam}
We have assessed the presence of stellar contamination in both the light curves and the transmission spectrum of \planet.
In \S\ref{SECTION-3:data-reduction}, we saw possible correlated noise in the white light curves, more prevalent in that of NRS1 than NRS2. 
This was seen in the Allan variance curves (Fig.~\ref{fig:allan}), and `bumps' were seen in the residuals of the detrended light curves (Fig.~\ref{fig:WLCs}).
Though such features may be induced by \textit{occultations} of stellar surface heterogeneities, within the transit chord, we saw no evidence for spot-crossing events when fitting the light curve.
At the transmission spectrum level, in \S\ref{section4:poseidon} we used the in-built functionality of \poseidon\ to simultaneously fit for a stellar contamination factor when performing atmospheric retrievals \citep{rackhamTransitLightSource2019}.
This test was limited to the scenario whereby the observed transmission spectrum is a combination of atmospheric chemical signal as well as contribution from \textit{unocculted} stellar surface heterogeneities.
While neither stellar contamination models were statistically preferred, it would be prudent to examine the plausibility of such a scenario.

Though TLSE contribution from F-type stars is predicted to be marginal \citep[][]{rackhamTransitLightSource2019}, there has been a range of activity signals reported in the literature.
For the F5-type star WASP-79, \citet{rathckeHSTPanCETProgram2021} saw evidence for faculae in their HST/STIS transmission spectrum, $\Delta T\sim500$\,K hotter than the photosphere, with a filling factor of 15\%, despite observed low photometric variability \citep[][]{sotzenTransmissionSpectroscopyWASP79b2019}.

Other examples include stars WASP-121 (F6 type star) and WASP-103 (F8), for which activity signals have varied between data and epochs.
For WASP-121, a faculae filling factor $f_\mathrm{fac}=0.08$ was derived from RV jitter \citep[][]{delrezWASP121HotJupiter2016}, while the HST/STIS transmission spectrum showed no signs of contamination \citep[][]{evansOpticalTransmissionSpectrum2018}.
WASP-103 displays little photometric variability \citep[][]{gillonWASP103NewPlanet2014}, but has a relatively high chromospheric emission $\log R'_\mathrm{HK} = -4.47$, which was noted to be higher than expected given the stellar age \citep[][]{staabSALTObservationsChromospheric2017}. 
\citet{kirkACCESSLRGBEASTSPrecise2021} then measured $f_\mathrm{fac}=0.22{\substack{+0.12\\-0.09}}$ with a temperature contrast $\Delta T=250$\,K from their optical + IR transmission spectrum of WASP-103\,b.

\mystar\ is thought to be a reasonably quiet, main-sequence F6V star.
\citet{naritaSpinOrbitAlignmentTrES42010} measured an RV variability of 20\,m/s, but this was only tentatively postulated to be due to stellar jitter \citep[see e.g.,][]{wrightRadialVelocityJitter2005}.
\mystar\ has been observed to display little chromospheric emission, with $\log R'_\mathrm{HK} = -5.11\pm0.15$ \citep{sozzettiNewSpectroscopicPhotometric2009}, and there have been no reports of activity-induced photometric variability \citep[e.g.,][]{knutsonDetectionTemperatureInversion2009,maciejewskiSearchPlanetsHot2023}. 

Though the temperature $T_\mathrm{het}$ we retrieved (\S\ref{section4:poseidon-chemeqm-activity}; see Table~\ref{table-app:retrieval-results-stellar-activity}) for \mystar\ is consistent with theory \citep[e.g.,][]{gondoinContributionSunlikeFaculae2008,rackhamTransitLightSource2019}, in comparison, $f_\mathrm{het}$ is uncharacteristically high, whilst also being poorly constrained, especially considering our target \mystar\ has a rotation rate around half that of WASP-79.
Indeed, assuming Sun-like heterogeneity location distributions, \citet{rackhamTransitLightSource2019} predicted a filling factor of $f_\mathrm{fac} = 0.01{\substack{+0.02\\-0.02}}$ for a F6V star.
On the other hand, though in theory there exists a degeneracy between the temperature of the heterogeneity and its filling factor \citep{pinhasRetrievalPlanetaryStellar2018}, the stellar contamination is easier to disentangle at the bluer wavelengths of the HST/STIS observations of \citet{rathckeHSTPanCETProgram2021} (shorter than $\sim 1\,$micron).

As explained in \S\ref{section4:poseidon}, since there is no additional evidence to support the models with stellar contamination included, the preferred models are those without.
Further, the key parameters presented in this work, namely the $\mathrm{C/O}$ and metallicity of the atmosphere of \planet, are not dependent on the choice of model.
From these JWST NIRSpec G395H observations, we thus conclude the atmospheric measurements of \planet\ regardless of the presence of stellar surface heterogeneities.
Future observations, for example at bluer wavelengths, will be required to constrain the level of stellar heterogeneity.

\section{Conclusions}
\label{SECTION-6:conclusions}
In this work we have presented the JWST/NIRSpec G395H $2.8-5.2$\,micron transmission spectrum of \planet.
Having implemented two independent reduction routines, we recover consistent transmission spectra.
We use two independent atmospheric retrieval analyses to characterise the atmosphere of \planet, presenting clear evidence for detections of H$_2$O, CO$_2$, and CO.  
We do not find sufficient evidence to suggest that cloud opacity impacts the transmission spectrum within the G395H bandpass, nor the need to account for stellar contamination, or an offset between detectors.
Though we do not detect other species at significance, we are able to place meaningful constraints on their abundances.
We place upper limits on the abundances of sulphur-bearing molecules \ch{H2S} and \ch{SO2}; while we see possible hints of \ch{H2S}, we rule out large abundances of \ch{SO2} (with a $2\sigma$ upper limit of $\log X_\mathrm{SO_2}<-6.97$).

Under chemical equilibrium, we retrieve consistent C/O across the reductions and retrieval setups. 
The C/O is sub-solar and likely lower than that of the host star, \mystar.
The retrieved metallicity is also likely to be sub-stellar, within the range $0.4-0.7\times$ solar.
The expectation from theoretical models of formation is either high C/O and low metallicity, or vice-versa. 
Both values being sub-stellar presents an interesting case with regard to formation scenarios.
We proposed two scenarios by which sub-stellar C/O and metallicity could be induced: the accretion of low C/O gas, or a combination of low-metallicity gas and low C/O solids accretion.
These scenarios could be distinguished with a tighter constraint on the sulphur abundance.

\planet\ is the second target in the BOWIE-ALIGN programme, in which there are four aligned and four misaligned targets.
We await the full sample before more significant links between atmospheric C/O, metallicity and formation can be inferred.

\section*{Acknowledgements}

This work is based on observations made with the NASA/ESA/CSA JWST. The data were
obtained from the Mikulski Archive for Space Telescopes at the Space Telescope Science
Institute, which is operated by the Association of Universities for Research in Astronomy,
Inc., under NASA contract NAS 5-03127 for JWST. 
These observations are associated with
program JWST-GO-3838. Support for program JWST-GO-3838 was provided by NASA
through a grant from the Space Telescope Science Institute, which is operated by the
Association of Universities for Research in Astronomy, Inc., under NASA contract NAS
5-03127. 

We thank the reviewer for their helpful comments, which helped to improve the quality of the manuscript.
RAB and AP thank the Royal Society for their support in the form of a University Research Fellowship. 
JK acknowledges financial support from Imperial College London through an Imperial College Research Fellowship grant.
HRW was funded by UK Research and Innovation (UKRI) under the UK government’s Horizon Europe funding guarantee [grant number EP/Y006313/1].
PJW acknowledges support from the UK Science and Technology Facilities Council (STFC) through consolidated grant ST/X001121/1.
NJM, DES, and MZ acknowledge support from a UK Reseach and Innovation (UKRI) Future Leaders Fellowship (Grant MR/T040866/1), a Science and Technology Facilities Funding Council Nucleus Award (Grant ST/T000082/1), and the Leverhulme Trust through a research project grant (RPG-2020-82).


\section*{Data Availability}

The data products associated with this manuscript can be found online
at Zenodo at \url{https://doi.org/10.5281/zenodo.15097695}.
We describe the data products resulting from our survey in \citet{kirkBOWIEALIGNJWSTComparative2024}.



\bibliographystyle{mnras}




\appendix
\section*{Appendices}

\section{Second \texttt{Tiberius} reduction}
\label{SECTION-7:appendix-tiberiusJK}

\begin{figure}
    \centering
    \includegraphics[width=0.98\linewidth]{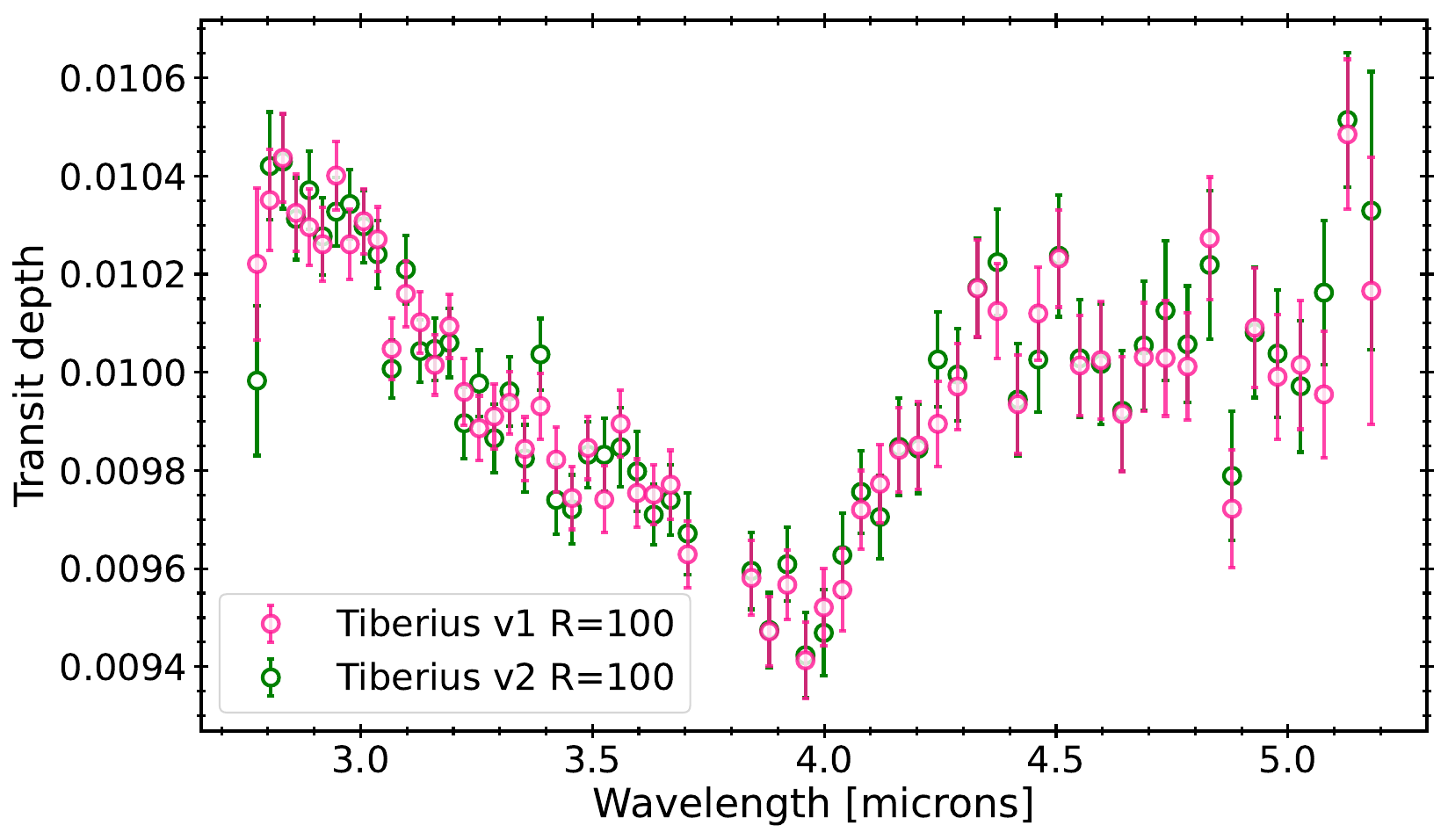}
    \caption{Comparison of the $R\simeq100$ transmission spectra from the two independent implementations of \tiberius{}. The spectrum in pink is the primary transmission spectrum analysed in this work (\S\ref{section:tiberius-LC-fitting}). }
    \label{fig-app:tiberius_comparison}
\end{figure}
As an additional independent check, we ran a second data reduction using the \texttt{Tiberius} package. This reduction was performed using the same \texttt{extraction\_input.txt} file, \texttt{jwst} pipeline version (1.8.2) and calibration reference files as used in the BOWIE-ALIGN WASP-15b study \citep{kirkBOWIEALIGNJWSTReveals2025}. We provide the list of used calibration reference files in our associated Zenodo repository. We performed this reduction to ensure one homogeneous reduction throughout the BOWIE-ALIGN programme. 

The key differences to the principal \texttt{Tiberius} reduction presented in \S\ref{SECTION-3:data-reduction} are the choice of aperture width (6 pixels in the principal vs.\ 8 pixels here) and the choice of \texttt{jwst} pipeline version (v1.13.4 vs.\ 1.8.2). 
Given the good agreement between both \texttt{Tiberius} reductions (Fig.~\ref{fig-app:tiberius_comparison}), this demonstrates that \planet's spectrum is insensitive to these differences in pipeline versions and calibration reference files. 
The light curve model was the same for this second reduction as for the principal reduction. 
However, for this second reduction, all light curves were fitted with Levenberg-Marquadt optimisation, rather than MCMC. 

Within this second reduction, we also explored the implications of choosing to fit the transit light curves with fixed quadratic limb darkening coefficients. Firstly, we refitted the data with a four parameter, non-linear law with coefficients fixed using the same stellar parameters, 3D \textsc{stagger} models and \texttt{ExoTiC-LD} software \citep{grantExoTiCLDThirtySeconds2024,magicStaggergridGrid3D2015} as done in the principal \texttt{Tiberius} reduction. 
This led to an almost identical spectrum as the fixed quadratic law (median difference of 3\,ppm). 
In a further test, we refit the light curves with both quadratic coefficients as free parameters and sampled the parameter space with MCMC. In this case, each wavelength bin in the transmission spectrum agreed to within $1\sigma$ of the fixed quadratic spectrum but with spectral uncertainties $1.53\times$ larger. Given these tests, we are confident that our transmission spectra are not biased by our choice of quadratic limb darkening with fixed coefficients.

\section{Astrophysical parameters from \texttt{Tiberius} light curves}
\label{SECTION-6:appendix-tiberius-wlc}

\begin{figure}

    \centering
    \includegraphics[width=0.95\linewidth]{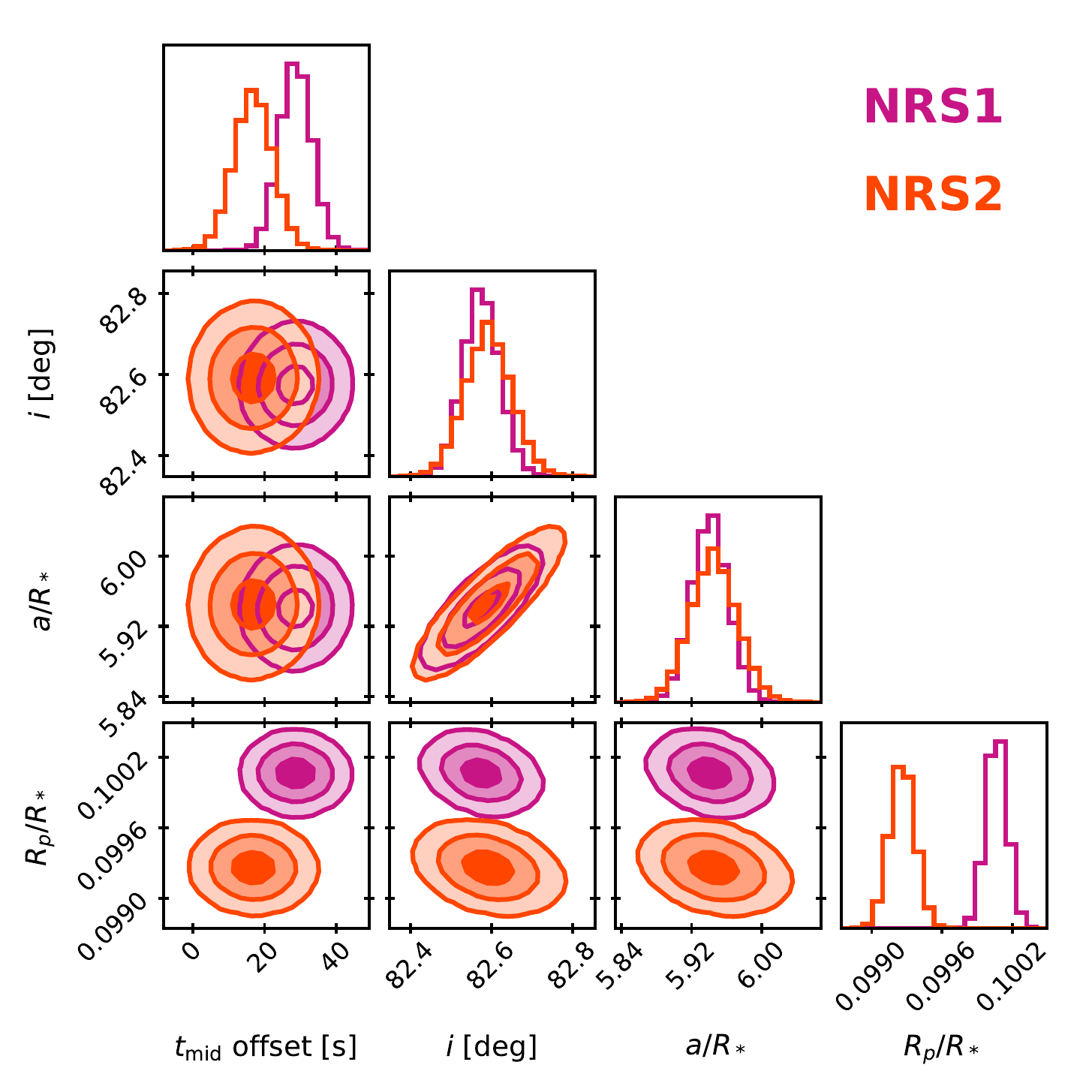}
    \caption{Posterior distribution of astrophysical parameters fitted in the \texttt{Tiberius} white light curves (each detector fitted independently).
    Contours ($1, 2, 3\sigma$) for NRS1 shown in dark pink and NRS2 in orange.}
    \label{fig-app:wlc-corner}
\end{figure}

In this appendix, we provide the posterior distributions for the individual \tiberius\ white light curves fits (Fig.~\ref{fig-app:wlc-corner}), described in \S\ref{section:tiberius-LC-fitting}.
The mean and 1$\sigma$ values are provided in Table~\ref{table:retrieved-params}.
We present the $t_\mathrm{mid}$ as an offset in seconds from the mid-transit time predicted by the propagated literature value (Table~\ref{table:retrieved-params}).
We note that there were two other free parameters in each of these fits: coefficients for the linear-in-time systematics model.

\section{Spectroscopic light curve analysis}
\label{SECTION-8:appendix-SLCs}

\begin{figure}
\centering
\includegraphics[width=0.98\linewidth]{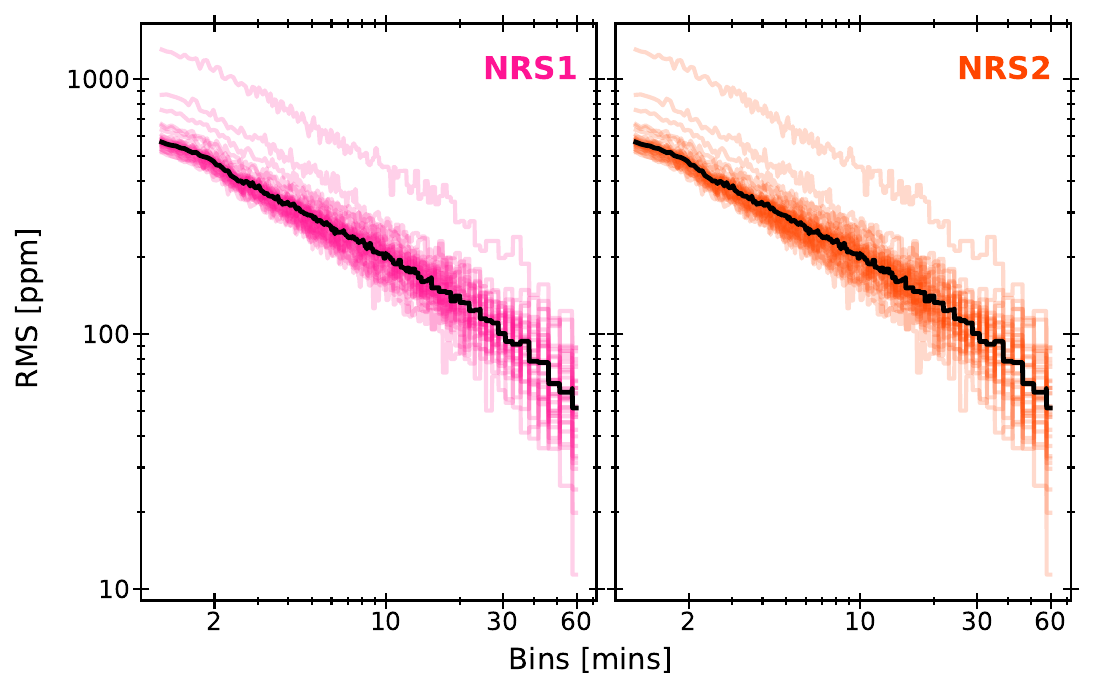}
\caption{The Allan variance of the \texttt{Tiberius} $R\simeq100$, detrended spectroscopic light curves. The medians across all spectroscopic bins are outlined in black.}
\label{fig-app:allan-slc}
\end{figure}

\begin{figure}

    \centering
    \includegraphics[width=0.98\linewidth]{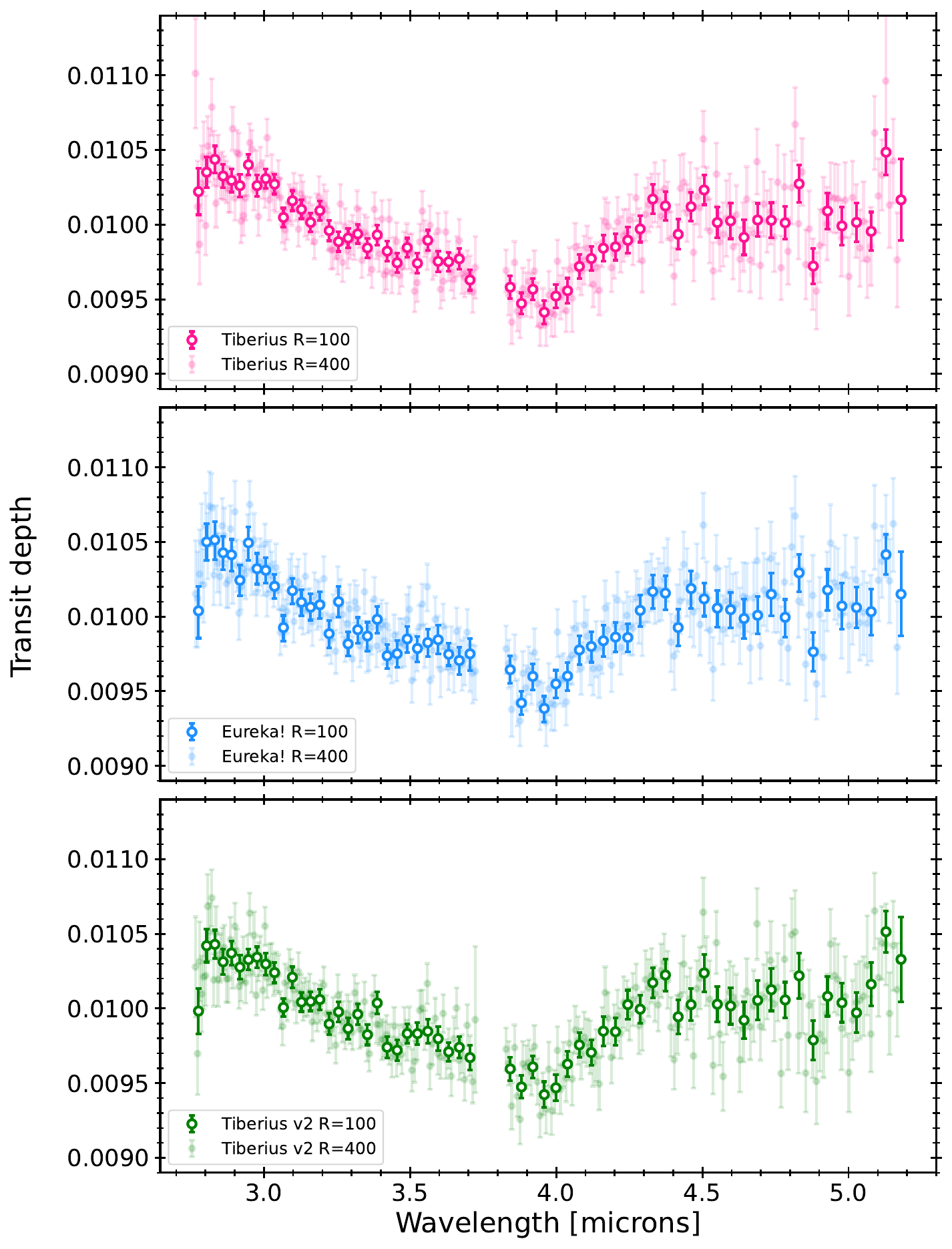}
    \caption{The $R\simeq100$ and $R\simeq400$ transmission spectra from the three independent reductions using \textbf{(top panel):} \tiberius\, \textbf{(middle):} \eureka\, and \textbf{(bottom):} \tiberius\ second reduction as detailed in Appendix~\ref{SECTION-7:appendix-tiberiusJK}.}
    
    \label{fig-app:R400}
\end{figure}

In \S\ref{SECTION-3:data-reduction}, we presented the $R\simeq100$ transmission spectrum of \planet.
The associated Allan variance plots for each of the detrended spectroscopic light curves are shown in Fig.~\ref{fig-app:allan-slc}; we found that a linear-in-time polynomial was sufficient to detrend these light curves.
We also presented results for the $R\simeq400$ transmission spectra, which are shown in Fig.~\ref{fig-app:R400} for each of the three reductions.

\section{Retrieval results}
\label{SECTION-9:appendix-retrieval-results}
\begin{figure}
\includegraphics[width=\columnwidth]{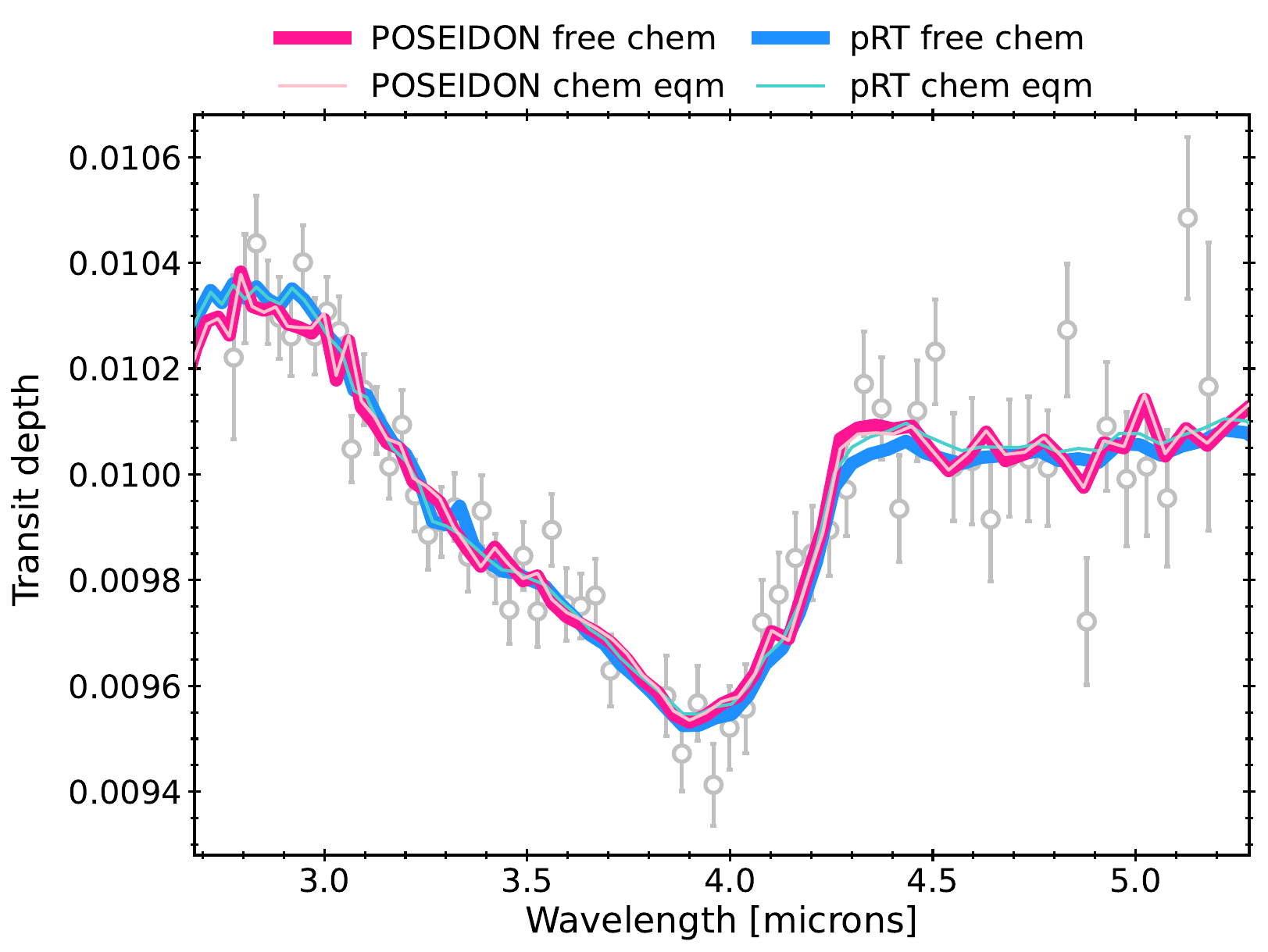}
\caption{The best-fit retrieved spectra from the \poseidon\ free and equilibrium chemistry retrievals on the \tiberius\ $R\simeq100$ transmission spectrum (data points in grey).
Also plotted are the best-fit spectra from \prt\ on the same spectrum.
The forward models have been binned to $R=100$.
}
\label{fig-app:all-retrieved-spec}
\end{figure}

\begin{figure*}
\includegraphics[width=\textwidth]{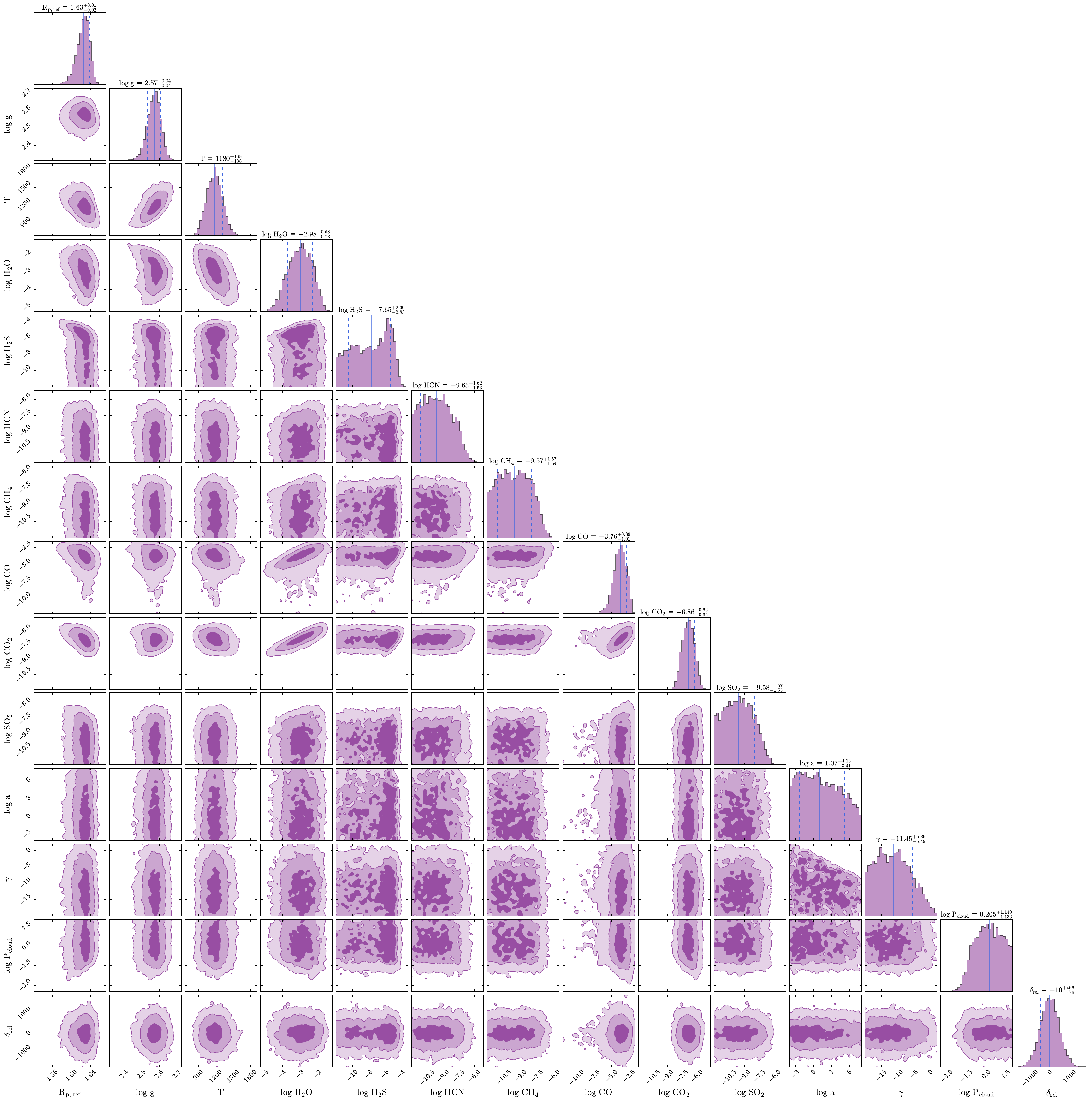}
\caption{Posterior distributions from the \poseidon\ free retrieval on the \texttt{Tiberius} $R\simeq100$ transmission spectrum (\S\ref{section4:poseidon-free}).}
\label{fig-app:free-corner}
\end{figure*}

\begin{figure*}
\centering
\includegraphics[width=\textwidth]{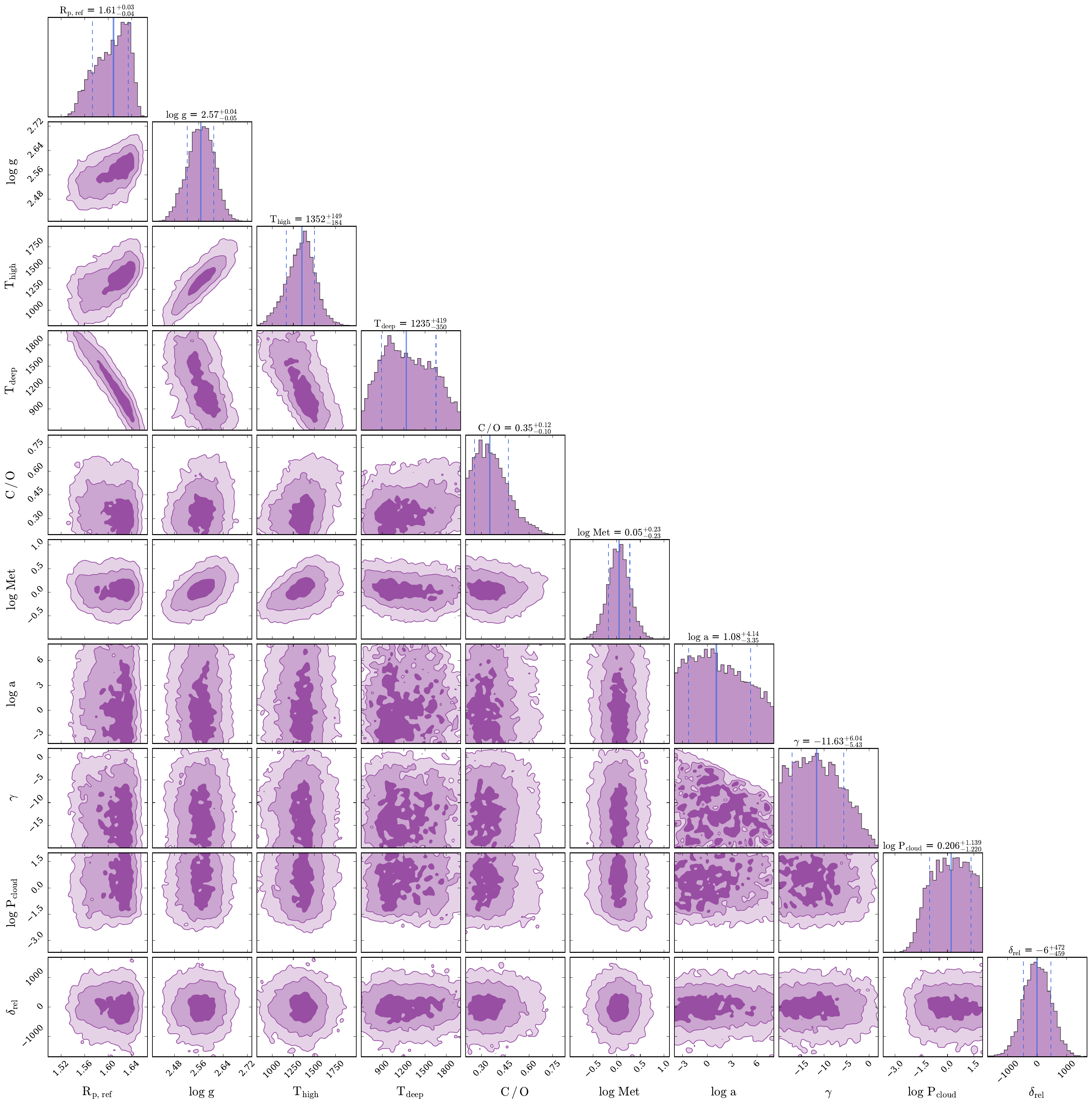}
\caption{Posterior distributions from the \poseidon\ equilibrium chemistry retrieval on the \texttt{Tiberius} $R\simeq100$ transmission spectrum (\S\ref{section4:poseidon-chemeqm}).}
\label{fig-app:chemeq-corner}
\end{figure*}

\begin{figure*}
    \centering
    \settototalheight{\dimen0}{\includegraphics[width=\textwidth]{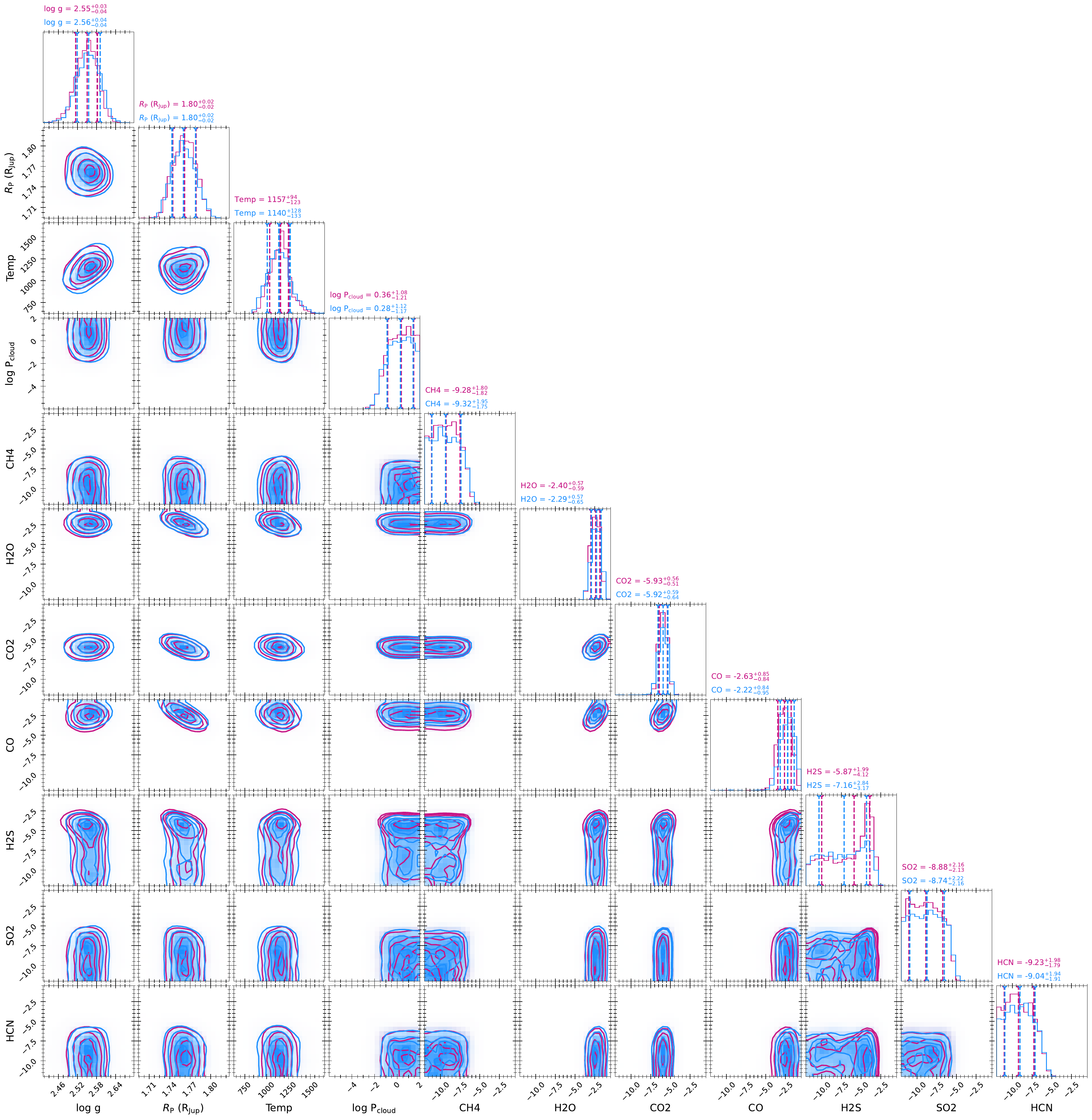}}
    \includegraphics[width=\textwidth]{prt_free_tres4_both_corner_Jan25.pdf}%
    \llap{\raisebox{\dimen0-5cm}{%
    \includegraphics[height=5cm]{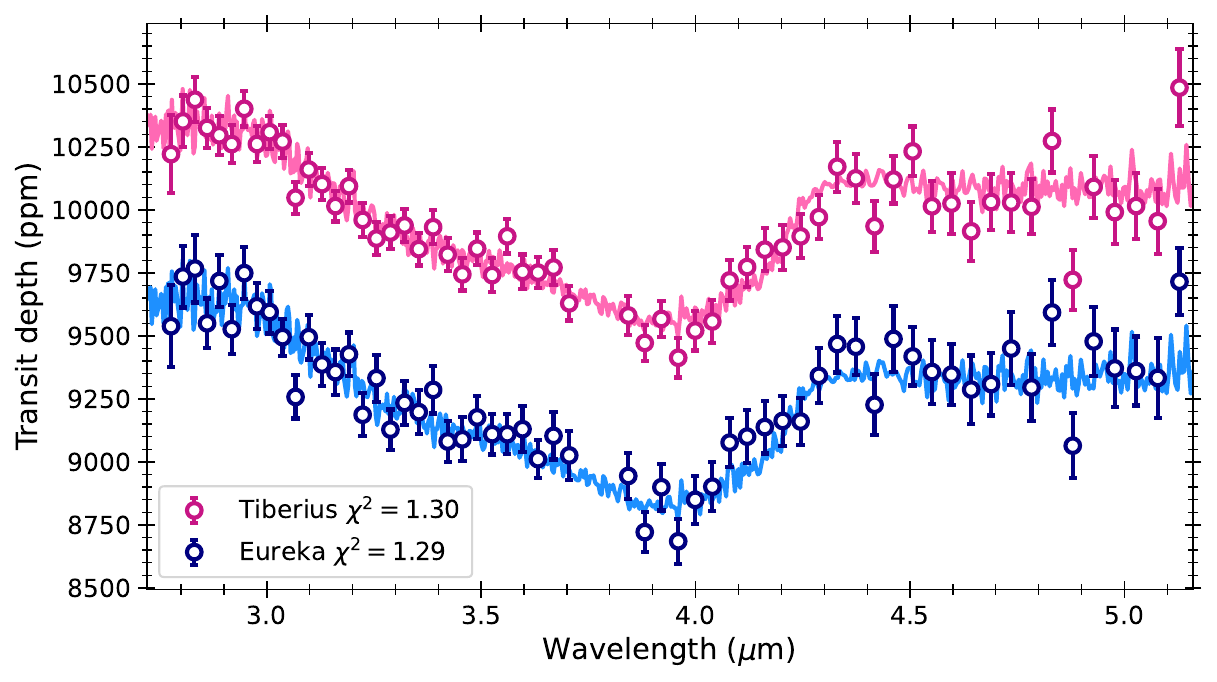}%
    }}
\caption{Posterior distributions from the \prt free chemistry retrieval (\S\ref{section:prt-retrievals}) on the $R\simeq100$ transmission spectra from \texttt{Tiberius} (pink) and \texttt{Eureka!} (blue). Species abundances are quoted in mass fraction, rather than mixing ratio. The best-fitting models are shown in the top right.}
\label{fig-app:pRT-free-corner}
\end{figure*}

\begin{figure}
    \centering
    \settototalheight{\dimen0}{\includegraphics[width=\textwidth]{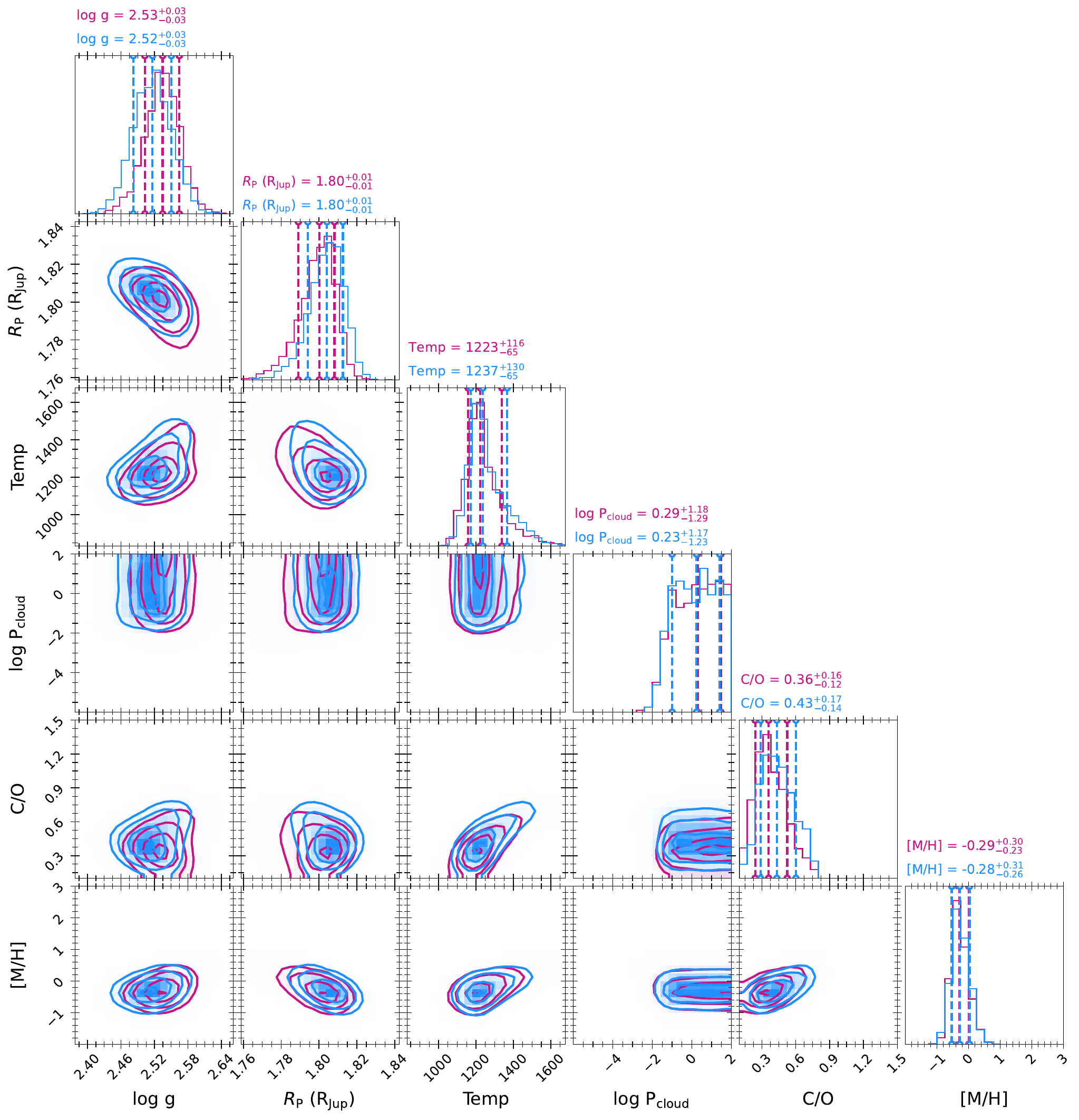}}
    \includegraphics[width=\textwidth]{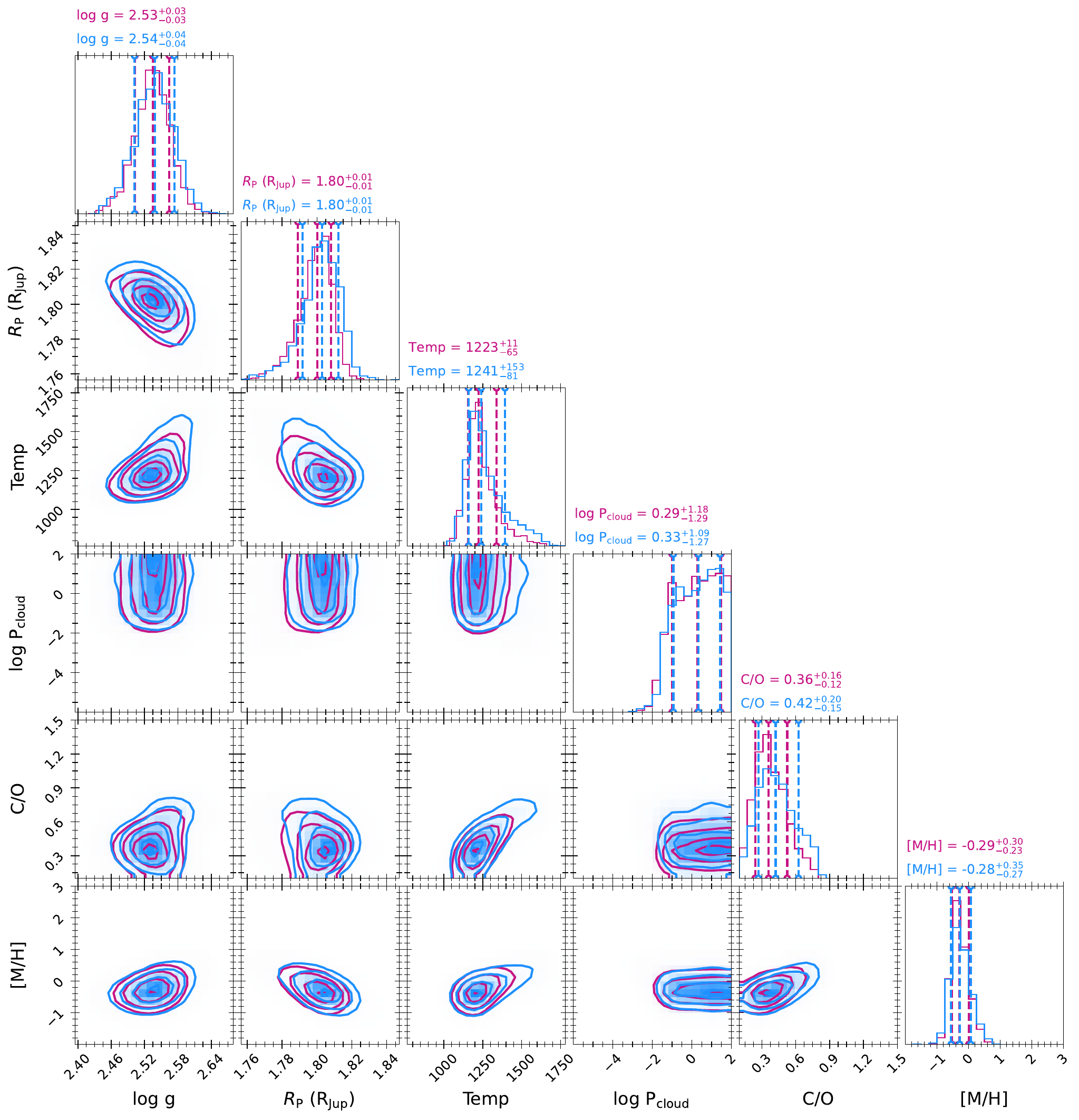}%
    \llap{\raisebox{\dimen0-5cm}{%
    \includegraphics[height=5cm]{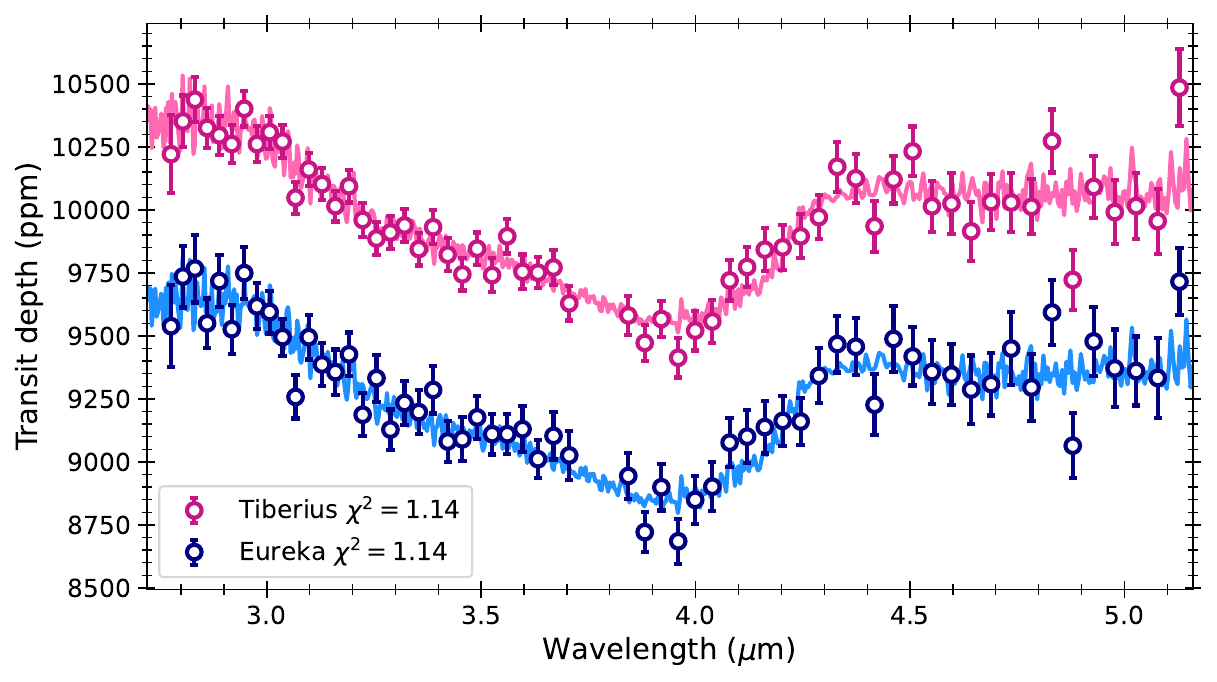}%
    }}
\caption{Posterior distributions from the \prt equilibrium chemistry retrieval (\S\ref{section:prt-retrievals}) on the $R\simeq100$ transmission spectra from \texttt{Tiberius} (pink) and \texttt{Eureka!} (blue). The best-fitting models are shown in the top right.}
\label{fig-app:pRT-chemeq-corner}
\end{figure}

\begin{figure*}
    \centering
    \settototalheight{\dimen0}{\includegraphics[width=\textwidth]{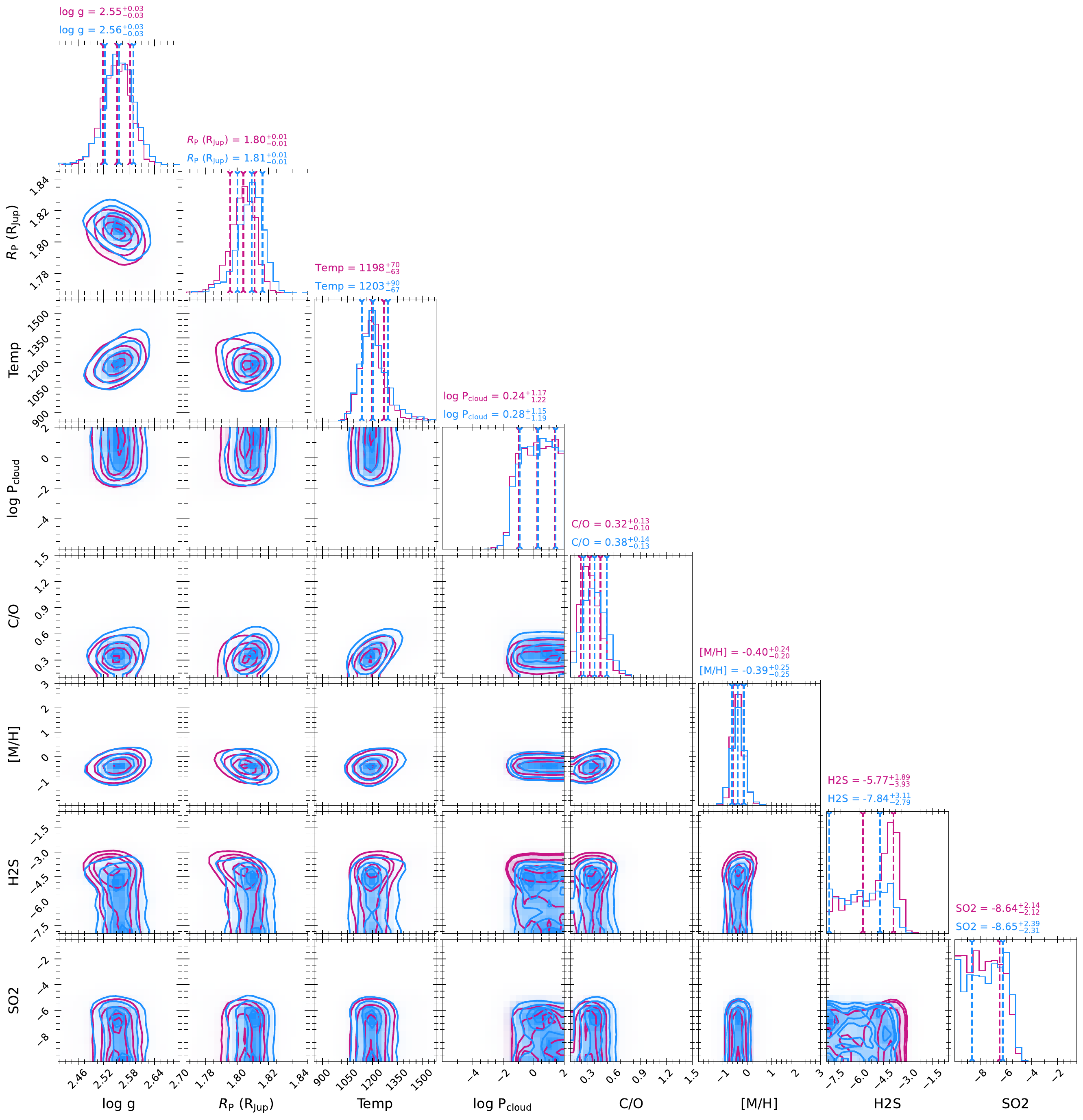}}
    \includegraphics[width=\textwidth]{prt_hybrid_tres4_both_corner_Jan25.pdf}%
    \llap{\raisebox{\dimen0-5cm}{%
    \includegraphics[height=5cm]{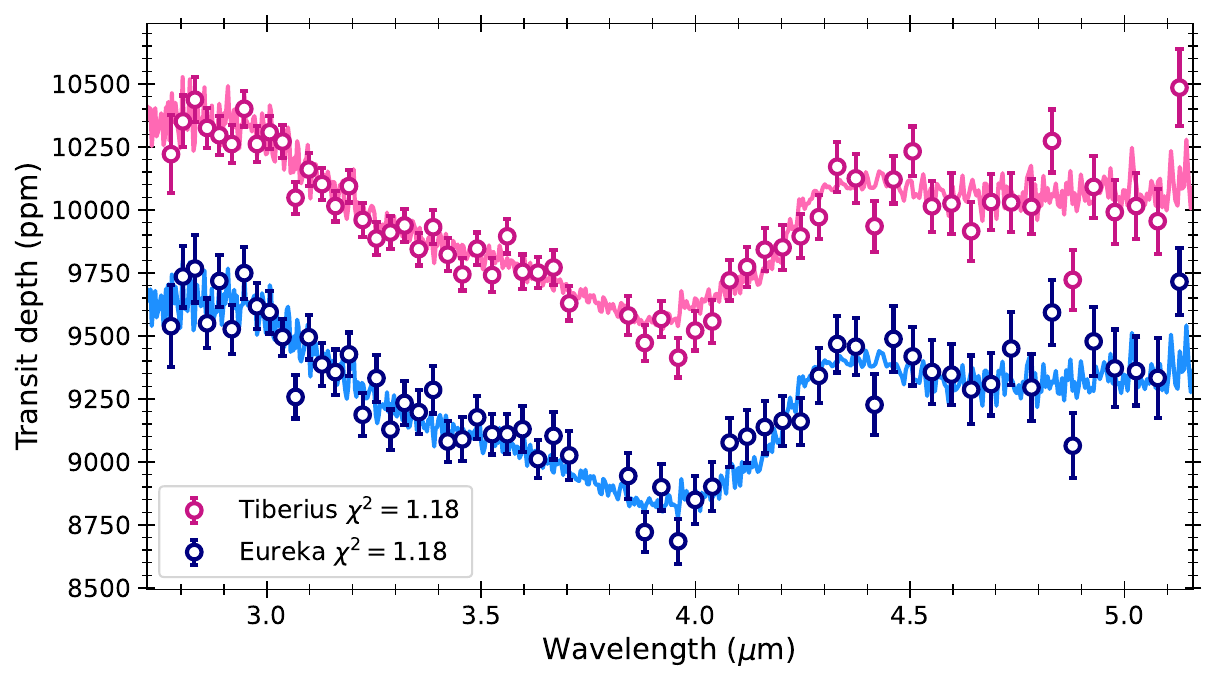}%
    }}
\caption{Posterior distributions from the \prt hybrid chemistry retrieval (\S\ref{section:prt-retrievals}) on the $R\simeq100$ transmission spectra from \texttt{Tiberius} (pink) and \texttt{Eureka!} (blue). Species abundances are quoted in mass fraction, rather than mixing ratio. The best-fitting models are shown in the top right.}
\label{fig-app:pRT-hybrid-corner}
\end{figure*}

Here we provide supplementary information for the retrieval tests outlined in \S\ref{SECTION-4:retrievals}.
A summary of the various priors used for the parameters of the retrievals are given in Tables~\ref{table-app:retrieval-priors} and \ref{table-app:retrieval-priors-stellar-activity}.
The resulting posterior distributions from the free and equilibrium chemistry \poseidon\ retrievals on the \tiberius\ $R\simeq100$ spectrum are shown in Fig.~\ref{fig-app:free-corner} and Fig.~\ref{fig-app:chemeq-corner} respectively.
We provide the posterior distributions and retrieved spectra from the \prt\ retrievals on the same transmission spectrum in Fig.~\ref{fig-app:pRT-free-corner}, Fig.~\ref{fig-app:pRT-chemeq-corner} and Fig.~\ref{fig-app:pRT-hybrid-corner}, and plot the best-fit \poseidon\ and \prt\ spectra together in Fig.~\ref{fig-app:all-retrieved-spec}.

\begin{figure}
    \centering
    \includegraphics[width=0.98\linewidth]{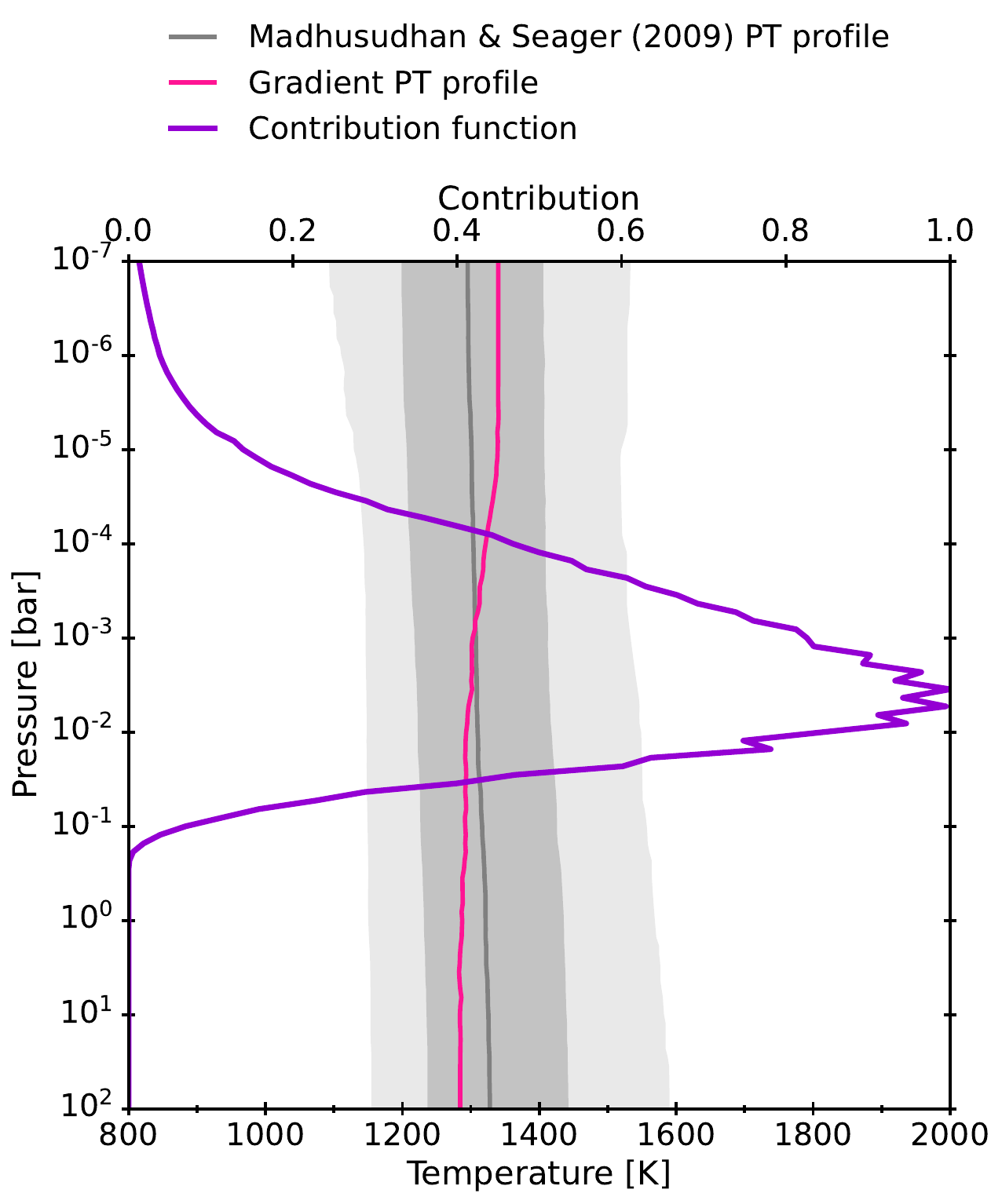}
    \caption{The median retrieved PT profile (mean and $1/2\sigma$ regions in grey) of the chemical equilibrium reference retrieval on the \tiberius\ $R\simeq100$ transmission spectrum, having implemented the more complex PT profile of \citet{Madhusudhan2009}.
    Overplotted in pink is the median retrieved gradient PT profile on the same spectrum (preferred), and in purple is the photometric contribution function across the full NIRSpec G395H bandpass.}
    \label{fig-app:madhu-PT}
\end{figure}

In \S\ref{section4:poseidon-chemeqm}, we decided to use a gradient PT profile for our \poseidon\ equilibrium chemistry retrievals. 
Here, we present in Fig.~\ref{fig-app:madhu-PT} the best-fitting PT profile. 
We compare the complex PT profile of \citet{Madhusudhan2009} to our simplified gradient PT profile.
We also plot the contribution function, using the functionality in \poseidon\ introduced by \citet{mullensImplementationAerosolMie2024}.
It can be seen that, in the region where the contribution function peaks (between $P=10^{-2}$ to 10$^{-3}$\,bar) we are probing similar temperatures. 
Given that the PT profile of \citet{Madhusudhan2009} is relatively isothermal throughout the region we are probing, and throughout the pressure ranges modelled, this justifies that a simplified PT profile is adequate to explain the observations. 
Furthermore, the location of the peak of the contribution function being higher than any of the cloud locations retrieved (see \S\ref{section4:poseidon}), aligns with the non-detections of any clouds within our observations.

Finally, in \S\ref{section4:poseidon-chemeqm-activity} we also explored the possibility of stellar heterogeneities, including a parameterisation in our \poseidon\ retrievals.
The results from the stellar contamination test \ref{eqm-contamination} in \S\ref{section4:poseidon-chemeqm-activity} are provided in Table~\ref{table-app:retrieval-results-stellar-activity}; the results from the other retrievals are given in Table~\ref{table:retrieval-results-species} and \ref{table:retrieval-results} in \S\ref{SECTION-4:retrievals}.

{
\renewcommand{\arraystretch}{1.3}
\begin{landscape}
\begin{table}
\caption{Priors used for the retrievals in \S\ref{SECTION-4:retrievals}. Wide uniform priors are used for all mixing ratios: $\log \mathrm{X}_i \sim \mathcal{U}[-12,-1]$.}
\begin{threeparttable}
    
    \begin{tabular}{l c c c cc c c c c c c} 
    \toprule
     Input spectrum & \# & $R_\mathrm{p,ref}\,[R_\mathrm{J}]$ & $\log g$\,[cgs] &  \multicolumn{2}{c}{$T$\,[K]$^\vdag$}  & $P_\mathrm{cloud}$\,[bar]  &  $\log a$ & $\gamma$ & $\mathrm{C/O}$ & $Z/Z_\odot$  & Offset\,[ppm] \\
     \midrule
     \poseidon & & & & & & & & &\\
     
     \textit{Free Chemistry} & \ref{free-baseline} & $\mathcal{U} [1.4,2.11]$ & $\mathcal{N} (2.60, 0.15) $ & 
     $\mathcal{U} [600,2000]$ & --& 
     $\mathcal{U}[10^{-7}, 10^2]$ & $\ddagger$ & $\ddagger$ & -- & -- &$\mathcal{N} (0,500)$ \\
     
     \textit{Equilibrium Chemistry} & \ref{eqm-baseline} & $\mathcal{U} [1.4,2.11]$ & $\mathcal{N} (2.60, 0.15) $ & $\mathcal{U} [600,2000]$ & $\mathcal{U} [600,2000]$ & $\mathcal{U}[10^{-7}, 10^2]$ & $\ddagger$ & $\ddagger$ & $\mathcal{U}[0.2, 2.0]$ &
     $\mathcal{U}[10^{-1}, 10^4]$ &$\mathcal{N} (0,500)$ \\
     
     \midrule 
     \prt & & & & & &   \\
     
     \textit{Free Chemistry} & \hyperlink{prt-free}{(IV)} & $\mathcal{U} [0.8,2.2]$ & $\mathcal{N} (2.56, 0.05) $ & 
     $\mathcal{U} [850,2850]$ & --& 
     $\mathcal{U}[10^{-6}, 10^2]$ & -- & -- & -- & -- & -- \\

     \textit{Equilibrium Chemistry} & \hyperlink{prt-chemeq}{(V)} & $\mathcal{U} [0.8,2.2]$ & $\mathcal{N} (2.56, 0.05) $ & $\mathcal{U} [500,3000]$ & -- & $\mathcal{U}[10^{-6}, 10^2]$ & -- & -- & $\mathcal{U}[0.1, 1.5]$ &
     $\mathcal{U}[10^{-2}, 10^3]$ & -- \\

     \textit{Hybrid Chemistry} &  \hyperlink{prt-hybrid}{(VI)} & $\mathcal{U} [0.8,2.2]$ & $\mathcal{N} (2.56, 0.05) $ & $\mathcal{U} [500,3000]$ & -- & $\mathcal{U}[10^{-6}, 10^2]$ & -- & -- & $\mathcal{U}[0.1, 1.5]$ &
     $\mathcal{U}[10^{-2}, 10^3]$ & -- \\

    \bottomrule
    \end{tabular}
\label{table-app:retrieval-priors}
\begin{tablenotes}
    \item[$\vdag$] If one value given, an isothermal temperature profile was used. If two values given, a gradient profile was used, and the values are ordered as $T_\mathrm{high}$ and $T_\mathrm{deep}$;
    \item[$\ddagger$] No user-specified prior, in which case \texttt{POSEIDON} applies a wide uniform prior.

\end{tablenotes}
\end{threeparttable}
\end{table}

\begin{table}
\caption{Priors used for the stellar activity retrieval in \S\ref{section4:poseidon-chemeqm-activity}. Other parameters follow the priors given in Table~\ref{table-app:retrieval-priors}.}

    \begin{tabular}{l c cc cc cc c c} 
    \toprule
     Input spectrum & \# & $f_\mathrm{het}$ & $T_\mathrm{het}$ & $\log_{g_\mathrm{het}}$ &  $T_\mathrm{phot}$  & $\log_{g_\mathrm{phot}}$  \\
     \midrule
     \poseidon & & & & & & \\
     
     \tiberius, $R=100$ & \ref{eqm-contamination} & $\mathcal{U}[0, 1]$ & $\mathcal{U}[0.8T_*, 1.2T_*]$ & $\mathcal{U}[3, 5]$  & $\mathcal{N}(T_*,\sigma_{T_*})$ & $\mathcal{N}(\log g_*,\sigma_{\log_{g_*}})$\\

    \bottomrule
    \end{tabular}
\label{table-app:retrieval-priors-stellar-activity}

\end{table}

\begin{table}
\caption{Results from the stellar contamination retrieval in \S\ref{section4:poseidon-chemeqm-activity}. Detailed of priors provided in Table~\ref{table-app:retrieval-priors-stellar-activity}.}

    \begin{tabular}{l c cc cc cc c c} 
    \toprule
     Input spectrum & \# & $f_\mathrm{het}$ & $T_\mathrm{het}$ & $\log_{g_\mathrm{het}}$ &  $T_\mathrm{phot}$  & $\log_{g_\mathrm{phot}}$  \\
     \midrule
     \poseidon & & & & & & & \\
     \tiberius, $R=100$ & \ref{eqm-contamination} & $0.22{\substack{+0.21\\-0.13}}$ & $6389{\substack{+388\\-313}}$ & $3.97{\substack{+0.63\\-0.58}}$ & $6306{\substack{+56\\-60}}$ & $4.09\pm0.03$\\

    \bottomrule
    \end{tabular}
\label{table-app:retrieval-results-stellar-activity}

\end{table}

\end{landscape}
}

\onecolumn

\bsp	
\label{lastpage}
\end{document}